\ifx\HCode\Undef
  \DocumentMetadata{
    lang        = en,
    pdfstandard = ua-2,
    tagging     = on
  }
\else
  \DocumentMetadata{}
\fi

\documentclass[manuscript]{acmart}

\AtBeginDocument{%
  }

\copyrightyear{2026}
\acmYear{2026}
\setcopyright{cc}
\setcctype{by}
\acmConference[ASSETS '26]{The 28th International ACM SIGACCESS Conference on Computers and Accessibility}{October 25--28, 2026}{Vila Nova de Gaia, Portugal}
\acmBooktitle{The 28th International ACM SIGACCESS Conference on Computers and Accessibility (ASSETS '26), October 25--28, 2026, Vila Nova de Gaia, Portugal}
\acmDOI{10.1145/3797867.3829070}
\acmISBN{979-8-4007-2521-0/2026/10}

\begin{document}

\title[HoldLight]{HoldLight: Making Indoor Climbing More Accessible for Blind and Low-Vision Climbers through Layered Route Preview and Real-Time Next-Hold Guidance}

\author{Chenyu Du}
\orcid{0009-0000-5192-5167}
\affiliation{%
  \institution{Xi'an Jiaotong-Liverpool University}
  \city{Suzhou}
  \country{China}
}
\email{chenyu.du23@student.xjtlu.edu.cn}

\author{Junyan He}
\orcid{0009-0006-1039-6004}
\affiliation{%
  \institution{Xi'an Jiaotong-Liverpool University}
  \city{Suzhou}
  \country{China}
}
\email{junyan.he23@student.xjtlu.edu.cn}

\author{Jiafei Sun}
\orcid{0009-0007-5252-3419}
\affiliation{%
  \institution{Xi'an Jiaotong-Liverpool University}
  \city{Suzhou}
  \country{China}
}
\email{jiafei.sun23@student.xjtlu.edu.cn}

\author{Yue Li}
\authornote{Corresponding author}
\orcid{0000-0003-3728-218X}

\affiliation{%
  \department{Academy of Artificial Intelligence and Advanced Technology}
  \institution{Xi'an Jiaotong-Liverpool University}
  \city{Suzhou}
  \state{Jiangsu}
  \country{China}
}

\affiliation{%
  \institution{University of Liverpool}
  \city{Liverpool}
  \country{United Kingdom}
}

\email{yue.li@xjtlu.edu.cn}
\email{yue.li@liverpool.ac.uk}

\renewcommand{\shortauthors}{Du et al.}

\begin{abstract}
Indoor climbing creates access barriers for blind and low-vision (BLV) climbers because route understanding and in-climb next-hold localization are typically visually mediated and socially dependent. We present \textit{HoldLight}, a workflow-centered accessibility system combining layered pre-climb route preview, concise real-time next-hold cueing, and explicit fallback under uncertainty. We grounded the design through formative inquiry with BLV climbers, refined the interaction via pilot testing, and conducted a controlled within-subjects evaluation comparing \textit{HoldLight} with a controlled on-demand human-support condition derived from current support practices. Findings provide initial evidence that structured support can improve pre-climb route understanding, support recovery from next-hold uncertainty, and reduce immediate human re-entry for routine route-information clarification under controlled conditions. \textit{HoldLight} contributes a bounded accessibility workflow, demonstrating where structured route-information is helpful, where explicit fallback is necessary, and how technical assistance can integrate with retained human safety oversight.
\end{abstract}




\begin{teaserfigure}
  \includegraphics[
  width=\textwidth,
  alt={A three-part storyboard illustration. Left: A climber with a mobile device scans a climbing wall. Center: The mobile interface displays a scanned wall with a target route, broken down into manageable segments. Right: The climber is on the wall, receiving auditory cues through an earpiece while progressing along the target route.}
  ]{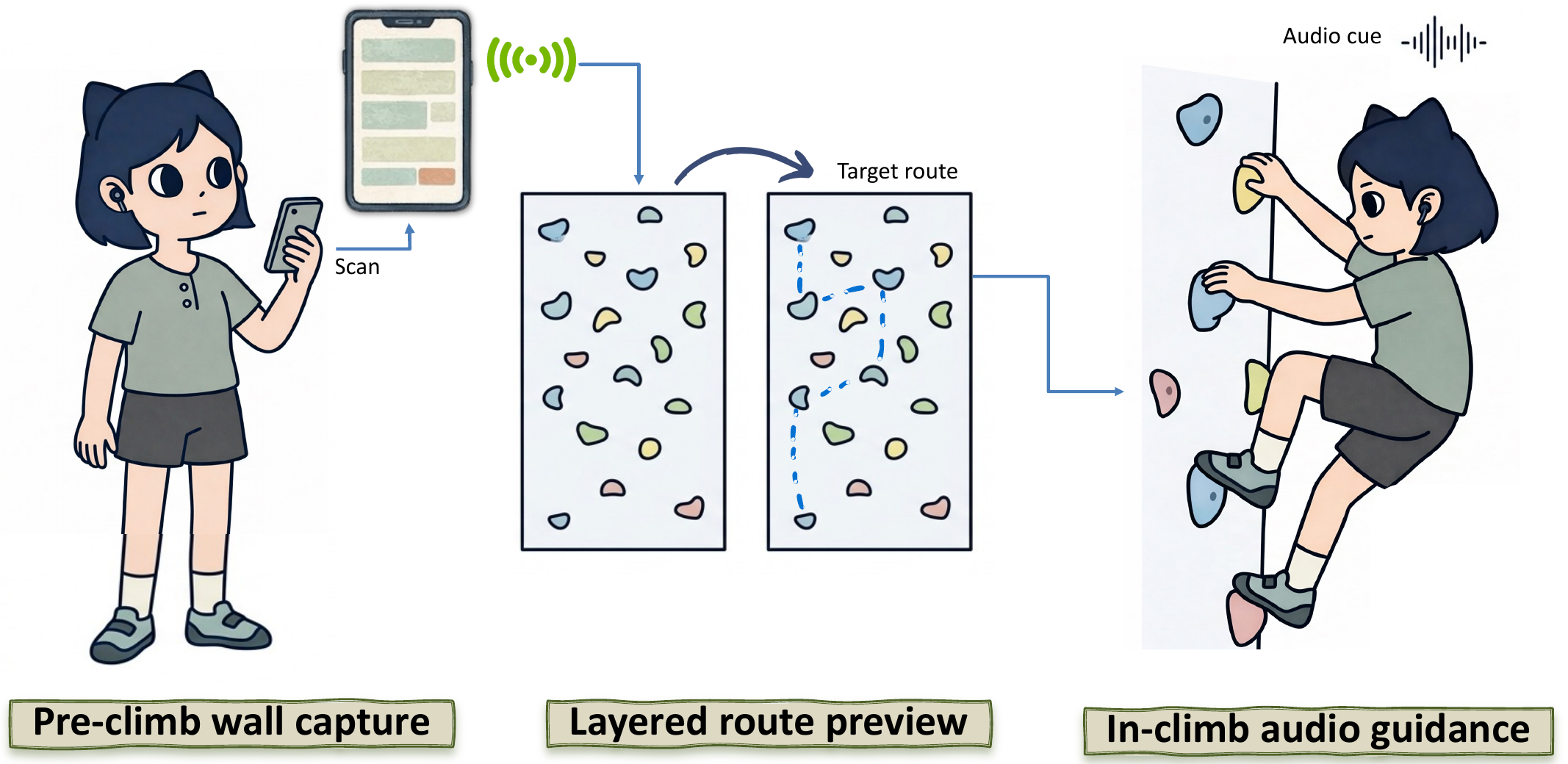}
  \caption{\textit{HoldLight} supports a bounded climbing-accessibility workflow for blind and low-vision climbers: the user captures the wall, receives a layered preview of the target route, and then climbs with concise next-hold audio guidance while human safety support remains available.}
  \Description{A three-part storyboard illustration. Left: A climber with a mobile device scans a climbing wall. Center: The mobile interface displays a scanned wall with a target route, broken down into manageable segments. Right: The climber is on the wall, receiving auditory cues through an earpiece while progressing along the target route.}
  \label{fig:teaser}
\end{teaserfigure}

\maketitle

\section{Introduction}

Indoor rock climbing conventionally depends on visually previewing a route before ascent and locating relevant holds as the climb unfolds \cite{Seifert2017RoutePreviewingClimbing}. Climbers typically inspect the intended route to form an initial plan, and must continually update this information while moving on the wall. For blind and low-vision (BLV) climbers, these visually mediated route-reading demands create access barriers. Prior climbing-accessibility systems have begun to address these barriers through localized audio feedback and vision-based sensory substitution \cite{Ilich2008MetaHolds,Richardson2022ClimbOVision}. Broader work on exercise and body-movement education suggests that support relying primarily on visual demonstration or verbal instruction alone can remain inaccessible or insufficient for BLV participants \cite{Rector2015ExerciseTechnologiesBLV,DeSilva2023BodyMovementEducationBLV,DeSilva2025SensingMovement}.

Systems for climbing have shown that climbing routes, holds, and movements can be supported through localized audio \cite{Ilich2008MetaHolds}, sensory substitution \cite{Richardson2022ClimbOVision}, projected visualization \cite{Kajastila2014AugmentedClimbing}, augmented climbing-wall interaction \cite{Kajastila2016AugmentedClimbingWall}, mobile AR authoring \cite{Daiber2013BouldAR}, and replay-based movement feedback \cite{Kosmalla2017ClimbVis}. Together, these works demonstrate the feasibility of technology-mediated climbing support, while also leaving open a route-information problem that is especially important for BLV climbers: how route information should be introduced before ascent, delivered concisely during bodily movement, corrected when uncertainty arises, and coordinated with human safety support.

Accessible climbing support draws on broader insights from accessible sports, exercise, navigation, and embodied guidance. Prior work has documented accessibility barriers in exercise and body-movement education for BLV people \cite{Rector2015ExerciseTechnologiesBLV,DeSilva2023BodyMovementEducationBLV,DeSilva2025SensingMovement}, shown the value of spatial preview before navigation \cite{Guerreiro2017VirtualNavigationBlindPeople,Aziz2022PlanningJourneyAudio}, and highlighted time-sensitive coordination and multi-sensory support during embodied activity \cite{Barbareschi2024SpeechIsSilver,DeSilva2025SensingMovement}. In indoor climbing, these requirements converge in a constrained form: BLV climbers need to build a usable route model before ascent, act on localized next-hold information under full-body exertion, and recover when route information becomes ambiguous or unavailable.

Route-information support for BLV climbers needs to be coordinated through interaction with a partner or guide rather than assumed to come entirely from a standalone system \cite{Kamikubo2020SupportStrategiesRemoteGuides,Vincenzi2021InterdependenceInAction}. Such support can be highly valuable, but it is also socially negotiated: asking for help may be awkward \cite{Teng2024HelpSupporters}, supporters vary in experience and verbal strategy \cite{Kamikubo2025BeyondOmakase}, and communication during movement may require sustained coordination \cite{Barbareschi2024SpeechIsSilver}. We therefore examine system support for a bounded part of routine route-information work, while keeping human guides central to safety, high-uncertainty judgment, and coaching.

In response, we present \textit{HoldLight}, an accessible climbing guidance system for BLV climbers organized around layered pre-climb route preview, concise real-time next-hold guidance, and explicit fallback and uncertainty handling. As illustrated in Fig.~\ref{fig:teaser}, \textit{HoldLight} structures route-information support across three linked moments: wall capture before climbing, structured preview for initial route understanding, and concise audio guidance during ascent. Computer vision and pose-based inference enable the system, while our contribution centers on how route information is organized and evaluated as an accessible workflow. We evaluate how route information can be structured so that BLV climbers can form an initial understanding of a route, request guidance when needed, maintain climbing rhythm, and remain in control when the system is uncertain or when human support is needed. To make this structured support usable in practice, the system also incorporates lightweight hold-color correction and route revision before interaction begins, reducing upstream ambiguities that could otherwise degrade route preview, next-hold guidance, and fallback behavior.

This paper addresses the following research questions:

\begin{enumerate}
\item[\textbf{RQ1}] What barriers, preferences, and success criteria do blind and low-vision climbers identify for accessible route-information support in indoor rock climbing?
\item[\textbf{RQ2}] How can an accessible climbing guidance system support blind and low-vision climbers' route-information access before and during indoor climbing?
\item[\textbf{RQ3a}] Under controlled indoor climbing conditions, how does \textit{HoldLight} compare with a controlled on-demand human-support condition in supporting route understanding and next-hold recovery?
\item[\textbf{RQ3b}] How do \textit{HoldLight} and the controlled on-demand human-support condition differ in perceived autonomy and immediate human re-entry for routine route-information clarification?
\end{enumerate}

Our contributions are threefold. First, we provide a formative account of barriers, preferences, and success criteria for accessible route-information support among BLV climbers. Second, we present \textit{HoldLight}, an accessible climbing guidance system organized around layered route preview, low-overload real-time cueing, and explicit fallback under uncertainty. Third, we provide initial mixed-method evidence from a controlled indoor evaluation, showing that this structured workflow improves some forms of route-information access and uncertainty recovery, supports greater perceived autonomy, and reduces immediate human re-entry for routine route-information clarification while leaving human safety support and high-uncertainty judgment clearly in place.

\section{Related Work}

\subsection{Accessible Climbing and Route-Information Technologies}

Prior work has explored several ways to make climbing routes, holds, and movements more accessible, interpretable, or trainable. For instance, MetaHolds supported hold discovery for blind climbers through head tracking and localized audio feedback \cite{Ilich2008MetaHolds}, while Climb-o-Vision used vision-based hold detection with electrotactile feedback to support access to climbing-hold information \cite{Richardson2022ClimbOVision}. These systems are closest to our work because they directly address non-visual access to climbing holds. However, their primary focus is on hold discovery, sensing, or feedback modality, and they do not evaluate a route-information workflow spanning pre-climb understanding, in-climb recovery, and explicit fallback.

Other climbing-support systems have focused on visual augmentation, route authoring, interaction with augmented climbing walls, and performance-oriented feedback. Prior work has explored projected route visualization \cite{Kajastila2014AugmentedClimbing}, augmented climbing-wall interaction \cite{Kajastila2016AugmentedClimbingWall}, mobile augmented-reality support for bouldering problem authoring and sharing \cite{Daiber2013BouldAR}, and replay or in-situ visualization of successful ascents for understanding climbing movements \cite{Kosmalla2017ClimbVis}. These systems establish important foundations for climbing-support technologies, showing that routes and climbing movements can be represented, augmented, or reviewed through interactive systems. However, these systems primarily address visual augmentation, collaborative route authoring, training, or performance reflection for climbers who can access visual feedback, whereas \textit{HoldLight} addresses a different access problem: how BLV climbers receive route information before ascent, act on concise next-hold cues under bodily load, and recover when system confidence or route interpretation becomes uncertain. \textit{HoldLight} therefore focuses on a BLV-centered route-information workflow for making routine route information available and recoverable before and during indoor climbing, complementing human support for coaching, expert route planning, and safety-critical judgment.

\subsection{Accessible Sports, Exercise, and Embodied Guidance for BLV Users}

Beyond climbing-specific systems, our work builds on accessibility research in exercise, sports, and body-movement education for BLV users. Prior research has shown that BLV people face persistent barriers to participation in exercise \cite{Rector2015ExerciseTechnologiesBLV}, sports, and other movement-based activities \cite{DeSilva2023BodyMovementEducationBLV}. Early work on exercise technologies identified opportunities and challenges around accessibility, independence, safety, and fit with real exercise practices \cite{Rector2013EyesFreeYoga,Rector2015ExerciseTechnologiesBLV,Rector2017EyesFreeYogaTACCESS,Malik2021TrainerLedAerobicExercise}. More recent work on body-movement education shows that BLV participation in activities such as dance, sports, and fitness involves not only access to instructions, but also bodily understanding, pacing, communication, and multi-sensory learning \cite{DeSilva2025SensingMovement,Jiang2025FromAudioDescriptionToMovement,DeSilva2023BodyMovementEducationBLV}. This line of work suggests that assistance for physical activity should be designed around how movement is actually carried out, rather than around abstract access to information alone.

Sports-specific accessibility work further shows that support must be adapted to the bodily and social structure of each activity. Prior work has examined BLV participation in running \cite{Abe2025ICanRunAtNight}, boxing \cite{Tsutsui2024LowVisionBoxing}, and VR boxing \cite{Kamath2025PunchPulse,Furtado2025VRBoxingBlindPeople}. Work on visually impaired running has studied both technological support for guided locomotion and the socially organized coordination between runners and guides \cite{Abe2025ICanRunAtNight,Barbareschi2024SpeechIsSilver,Yu2025RunPacer}. Related work has also considered directional awareness in running support \cite{AlZayer2016DroneGuideBlindRunners}, while accessible boxing and VR boxing research highlights participatory design, multimodal feedback, exertion, orientation, and autonomy in physically active settings \cite{Kamath2025PunchPulse,Furtado2025VRBoxingBlindPeople,Tsutsui2024LowVisionBoxing}. These studies are relevant because they treat accessibility support as part of dynamic embodied activity. However, climbing differs from running, boxing, and structured exercise instruction systems because route structure and action guidance are tightly coupled: climbers must form a route model before ascent and then repeatedly act on localized hold information while maintaining balance, rhythm, and bodily effort on the wall \cite{Daiber2013BouldAR,Kajastila2016AugmentedClimbingWall}.

Work on accessible navigation and journey planning also informs our focus on preview before movement. Prior research shows the value of previewing spatial structure before travel or navigation \cite{Guerreiro2017VirtualNavigationBlindPeople,Aziz2022PlanningJourneyAudio}. However, indoor climbing differs from planar navigation because route preview must be translated into hold-by-hold vertical action under bodily load. In this sense, climbing is neither equivalent to continuous guided locomotion \cite{Abe2025ICanRunAtNight,Barbareschi2024SpeechIsSilver} nor to exercise systems with fixed prompts \cite{Rector2013EyesFreeYoga,Rector2017EyesFreeYogaTACCESS,Malik2021TrainerLedAerobicExercise}. Our work therefore extends accessible sports and embodied guidance research into a route-structured climbing context, where preview, localized target information, pacing, and recovery must remain aligned during ascent.

\subsection{Workflow-Centered AI and Context-Sensitive Accessibility Support}

A growing body of accessibility research has explored how AI and interactive systems can support BLV users in understanding complex content, navigating task structures, and receiving context-sensitive assistance \cite{Zhang2025A11yShape,Anderer2025LectureVideosAccessible,Kim2023ChartQAforBLV,Wang2023TouchPilot,Peng2021Slidecho}. 
% Recent ASSETS studies such as A11yShape \cite{Zhang2025A11yShape} and  LectureAssistant \cite{Anderer2025LectureVideosAccessible} are especially instructive. 
Although these systems operate in domains very different from climbing, they show how AI support can be integrated into user-centered workflows for interpretation, navigation, interaction, and verification. In particular, Zhang et al. \cite{Zhang2025A11yShape} present A11yShape, an AI-assisted 3D modeling system that supports BLV programmers through linked representations, hierarchical navigation, and iterative verification. Anderer et al. \cite{Anderer2025LectureVideosAccessible} introduce LectureAssistant, which enables BLV students to access lecture videos through AI-assisted navigation and visual question answering. In both cases, users are not merely given a single AI-generated output; instead, assistance is structured through representations, navigation, question asking, or iterative access to information. Their relevance to our work therefore lies less in the application domain, but in their workflow-centered framing of accessibility support. We adapt them to a physically ongoing climbing task where support must remain brief, actionable, and compatible with bodily pacing.

This workflow-centered view is also reflected in recent systems for accessible content interaction and personalization \cite{Sechayk2025VeasyGuide,Natalie2024AudioDescriptionCustomization,Ning2024SPICA,Cheema2025DescribeNow}. For example, Sechayk et al. \cite{Sechayk2025VeasyGuide} present VeasyGuide, a system that supports low-vision learners in locating and interpreting instructor actions in presentation videos through personalized visual guidance. More broadly, Jiang et al. \cite{Jiang2024ContextDependentVideoAccessibility} show that BLV users' video accessibility preferences can vary substantially across viewing scenarios, while Stangl et al. \cite{Stangl2021OneSizeFitsAllImageDescriptions} argue against one-size-fits-all accessibility assumptions in visual content access. These systems suggest that accessibility support should be organized around meaningful user tasks, adapt to context rather than assume one universal mode of support, and preserve user control, personalization, and smooth interaction flow. However, most work focuses on content access and structured exploration \cite{Ning2024SPICA,Peng2021Slidecho}, or personalized presentation \cite{Natalie2024AudioDescriptionCustomization,Cheema2025DescribeNow,Sechayk2025VeasyGuide}, whereas \textit{HoldLight} applies a similar workflow-centered stance to an embodied climbing activity where information must be acted on during movement and remain recoverable under bodily load.

\subsection{Real-Time Cueing, Fallback, and Human Collaboration}

A final line of relevant work concerns how spatial, auditory, haptic, and embodied guidance should be delivered in real time. Prior work has examined real-time and spatial guidance through auditory, haptic, and wearable systems, including track navigation feedback \cite{Rector2018AuralHapticTrack}, continuous tactile path-following guidance \cite{Liu2021TactileCompass}, smartglasses-based wayfinding \cite{Zhao2020WayfindingSmartglassesLowVision}, audio-haptic finger guidance for document reading \cite{Stearns2016HapticAuditoryDirectionalGuidance}, and audio-haptic guidance for tactile graphics exploration \cite{Chase2020PantoGuide}. Rahman et al. \cite{Rahman2023TakeMyHand} present Take My Hand, showing that automated hand-based spatial guidance can help visually impaired users reach targets on a surface. This literature suggests that guidance design depends not only on what information is conveyed, but also on when it is delivered, through which modality, and how much interpretation burden it imposes on the user \cite{Rector2018AuralHapticTrack,Zhao2020WayfindingSmartglassesLowVision,Stearns2016HapticAuditoryDirectionalGuidance,Barbareschi2024SpeechIsSilver}. These insights are directly relevant to climbing, where an overlong or poorly timed instruction may disrupt movement, and where cues must support not only target localization but also upward progression, balance management, and rhythm preservation.

Real-time accessibility systems must also account for uncertainty, error, and the limits of automation. Prior work on AI-enabled visual assistance \cite{Alharbi2024MisfittingWithAI,Hong2024ObjectRecognitionErrors} highlights persistent errors and the difficulty of non-visually verifying AI output. Work on remote sighted assistance and human--AI collaboration shows that human support remains important in real-world access tasks, particularly when interpretation, judgment, or breakdown repair is needed \cite{Lee2022HumanAICollaborationRSA,Kamikubo2020SupportStrategiesRemoteGuides,Lee2020EmergingProfessionalPracticeRSA}. Studies of guided running \cite{Barbareschi2024SpeechIsSilver}, sighted guiding \cite{Vincenzi2021InterdependenceInAction}, and remote sighted assistance \cite{Lee2022HumanAICollaborationRSA,Kamikubo2020SupportStrategiesRemoteGuides} further show that support often involves coordinated action over time rather than fixed instruction delivery alone. This body of work motivates treating interdependence, fallback, and human collaboration as integral, designed components of accessibility support \cite{Bennett2018InterdependenceFrame,Alharbi2024MisfittingWithAI,Lee2022HumanAICollaborationRSA}.

In climbing, uncertainty may arise from missed holds, lost tracking, ambiguous cues, dense same-color hold clusters, or elevated safety concern. Drawing on prior work on accessible AI error handling \cite{Alharbi2024MisfittingWithAI}, mixed-initiative interaction \cite{Peng2025Morae}, user-controlled information delivery \cite{Cheema2025DescribeNow}, and support for requesting human assistance \cite{Teng2024HelpSupporters}, we argue that a usable climbing-support workflow should make uncertainty legible and provide clear recovery paths. Accordingly, \textit{HoldLight} operationalizes fallback as an explicit interaction layer: users can repeat information, request more detail, pause, or escalate to human help, while human safety oversight and high-uncertainty judgment remain outside the system's automation boundary.

\section{Phase A: Formative Inquiry}

\begin{figure*}[h]
  \centering
  \includegraphics[
    width=\textwidth,
    alt={A three-phase research workflow. Phase A uses formative inquiry with blind and low-vision participants and thematic analysis to derive four design goals. Phase B uses pilot testing with blind and low-vision climbers to refine preview structure, cue vocabulary, and fallback controls. Phase C compares HoldLight with a controlled on-demand human-support condition across route understanding, uncertainty recovery, perceived autonomy, and immediate human re-entry for routine route-information clarification.}
  ]{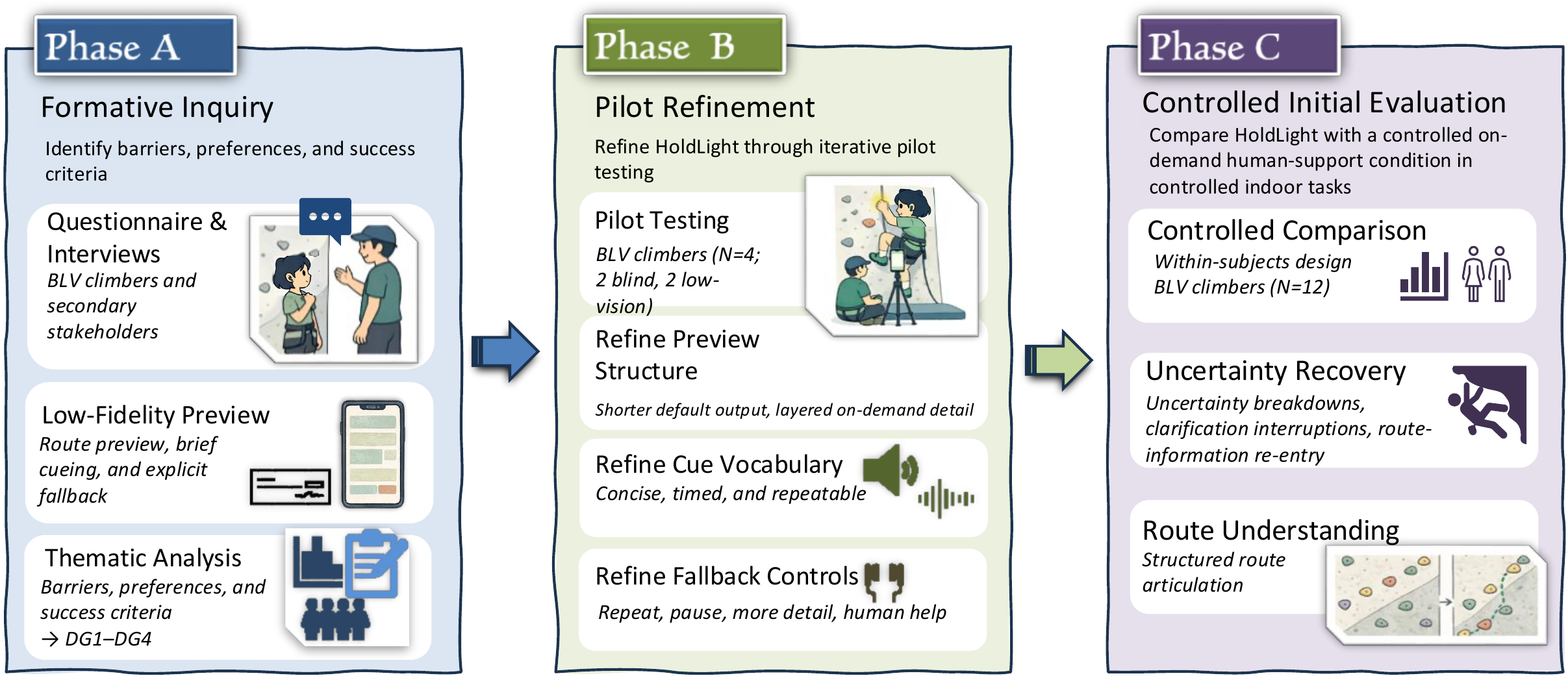}
  \caption{Overview of the three-phase human-centered research process. Phase A identifies barriers and derives four design goals (DG1--DG4), Phase B refines preview, cueing, and fallback interaction details, and Phase C compares \textit{HoldLight} with a controlled on-demand human-support condition.}
  \Description{A three-phase research workflow. Phase A uses formative inquiry with blind and low-vision participants and thematic analysis to derive four design goals. Phase B uses pilot testing with blind and low-vision climbers to refine preview structure, cue vocabulary, and fallback controls. Phase C compares HoldLight with a controlled on-demand human-support condition across route understanding, uncertainty recovery, perceived autonomy, and immediate human re-entry for routine route-information clarification.}
  \label{fig:methodology}
\end{figure*}

Fig.~\ref{fig:methodology} provides an overview of the three-phase study design.
To address \textbf{RQ1}, we conducted a formative inquiry to understand how blind and low-vision (BLV) climbers currently preview routes, locate the next hold during climbing, manage uncertainty, and work with human support in indoor climbing contexts. The purpose of this phase was not to evaluate a finished system, but to identify barriers, preferences, success criteria, and unacceptable support experiences that could ground subsequent design. In addition to perspectives from BLV climbers, we included a small number of secondary stakeholders: sighted guides and coaches with experience supporting BLV climbers or with professional climbing instruction. Their accounts helped contextualize current support practices, communication breakdowns, and the boundary between routinizable information support and human judgment, while design implications were grounded primarily in BLV participants' accounts.

\subsection{Participants}

Participant characteristics across the formative and controlled evaluation phases are summarized in Table~\ref{tab:participant_characteristics}. Phase A involved two complementary participant groups. First, eleven blind and low-vision (BLV) participants completed a short background questionnaire, of whom four additionally took part in in-depth semi-structured interviews. Second, five secondary stakeholders with direct experience supporting visually impaired climbers in practice---three sighted guides and two coaches---participated in separate semi-structured interviews. Sighted guides are people who are acquainted with the BLV climbers they support and usually accompany them to provide real-time guidance; coaches are climbing-gym staff, usually professional climbers, who are not necessarily familiar with the BLV climbers they support or with their specific support needs. Design implications were grounded primarily in BLV participant accounts, while stakeholder interviews were used to contextualize current support practices, timing constraints, supporter workload, and the boundary between routinizable information support and human judgment.

\begin{table*}[h]
\caption{Participant characteristics across the formative and controlled evaluation phases. In Phase A, the 4 BLV interviewees were a subset of the 11 BLV questionnaire respondents rather than an additional BLV sample.}
\label{tab:participant_characteristics}
\centering
\footnotesize
\setlength{\tabcolsep}{4pt}
\renewcommand{\arraystretch}{1.05}

\tagpdfsetup{table/header-rows={1}}
\begin{tabular}{p{0.27\textwidth} p{0.20\textwidth} p{0.16\textwidth} p{0.30\textwidth}}
\toprule
\textbf{Research method, phase, and sample size ($N$)} & \textbf{Vision status / role} & \textbf{Climbing experience} & \textbf{Key insights / structure} \\
\midrule

Questionnaire for BLV climbers in Phase A ($N=11$)
& Blind (including minimal light / contour perception): 6; Low vision: 5
& $<$6 months: 3; 6--12 months: 3; 1--3 years: 5
& Most BLV climbers rely heavily on a sighted guide or coach for route preview and/or in-climb prompting. \\

Semi-structured interview with BLV climbers in Phase A ($N=4$)
& Blind: 2; Low vision: 2
& $<$6 months: 1; 6--12 months: 1; 1--3 years: 2
& In-depth semi-structured interviews on route preview, next-hold guidance, uncertainty handling, autonomy, and unacceptable support experiences. \\

Semi-structured interview with sighted guides and coaches in Phase A ($N=5$)
& Sighted guides: 3; Coaches: 2
& $>$3 years: 5
%& Guides: 3; Coaches: 2
%& Not applicable
& Semi-structured interviews on current support practices, communication breakdowns, supporter workload, and the boundary between routinizable information support and human judgment. \\

System evaluation with BLV climbers in Phase C ($N=12$)
& Blind (including minimal light / contour perception): 6; Low vision: 6
& $<$6 months: 3; 6--12 months: 3; 1--3 years: 6
& Controlled within-subjects evaluation comparing \textit{HoldLight} with a controlled on-demand human-support condition derived from current support practices. \\

\bottomrule
\end{tabular}
\end{table*}

\subsection{Procedure}

Phase A combined a short accessible background questionnaire with semi-structured online interviews. The questionnaire was designed to gather basic contextual information about participants’ visual condition, climbing experience, current support arrangements, preferred information channels during movement, perceived current barriers, and broad attitudes toward technology-assisted climbing support. Its role was not to serve as a stand-alone outcome measure, but to establish participant context, invite voluntary interview participation, and help focus later interviews on issues such as route preview, real-time prompting, pacing, uncertainty, and acceptable boundaries of assistance.

The four BLV semi-structured interviews were conducted online and lasted approximately one hour each. The interviews focused on participants’ current climbing practices, including how they typically understand a route before climbing, how they locate the next hold while climbing, when and why they seek human help, and which kinds of support they experience as helpful, disruptive, or unacceptable. We also probed concrete positive and negative incidents to move beyond general preferences and capture what had actually happened in practice, for example, times when route explanation failed, when real-time prompting supported a move effectively, or when poorly timed instructions interrupted rhythm or confidence. Toward the end of the interview, we introduced a low-fidelity support concept involving route preview, brief in-climb prompting, and explicit fallback controls in order to explore design opportunities without treating the session as a prototype evaluation.

The five secondary stakeholder interviews were also conducted online and lasted approximately one hour on average. These interviews focused on current support ecology rather than on speaking for BLV climbers. Specifically, we asked sighted guides and coaches how they typically explain routes before a climb, how they decide when to speak during a climb, what kinds of spatial wording or timing problems recur in practice, which moments are most cognitively demanding for supporters, and how they would draw the boundary between stable, repeatable support that a system might eventually help provide and high-risk or situational judgment that should remain human-led.

All participant-facing phases of the study were covered by institutional ethics approval. Participants received accessible study information and provided informed consent before taking part. For online interviews, participants could skip questions or stop participation at any time.

\subsection{Analysis}

We analyzed the formative materials using a hybrid deductive--inductive thematic approach \cite{Fereday2006DemonstratingRigor,Braun2006UsingThematicAnalysis}. Our initial deductive lenses were derived from the scope of \textbf{RQ1} and from the interview and questionnaire design, namely: route preview before climbing, next-hold guidance during climbing, pacing and rhythm, uncertainty and safety, autonomy and enjoyment, and the role of human support. At the same time, we remained open to emergent patterns that were not fully captured by these lenses, such as concerns about social burden, reluctance to over-rely on others, and differences between wanting help and wanting to remain in control.

In the first stage of analysis, we reviewed questionnaire responses to identify broad participant tendencies relevant to later design work, such as the degree of dependence on others for route preview, concern about information overload, willingness to receive audio support during movement, and concern about system latency, inaccuracy, or overconfidence. In the second stage, we examined the interview materials to identify repeated barriers, preferences, and criteria for effective support. Rather than fragmenting the data into many small themes, we focused on a small set of strong cross-cutting themes that could directly inform system design.

Two authors reviewed the formative materials using the deductive lenses above, compared candidate codes, and reconciled disagreements through discussion. Candidate themes were retained when they recurred across BLV accounts and directly informed design decisions concerning route preview, in-climb cueing, uncertainty recovery, or human-support boundaries. Themes that were isolated, outside route-information support, or not actionable within the system scope were treated as contextual observations rather than retained design themes. Guide and coach interviews were used to contextualize support practices and safety boundaries, not to override BLV participant accounts.

\subsection{Findings}

Examining the interviews with the BLV climbers and secondary stakeholders, we found four themes that consistently captured the key barriers, preferences, and criteria for effective accessible climbing guidance: (1) route preview was fragmented and often socially dependent, (2) real-time guidance was useful but could easily overload timing and action, (3) uncertainty needed explicit management, and (4) effective support had to preserve pace, enjoyment, and autonomy while avoiding over-instruction.

%\subsubsection{Route Preview Is Fragmented and Socially Dependent}
\subsubsection{Route preview is fragmented and socially dependent}

Participants suggested that understanding a route before leaving the ground was both important and difficult. Several BLV climbers described beginning a climb with only a partial understanding of where the route went. As one participant (P3) noted, \textit{``I usually just nod when they explain the sequence on the ground, but honestly, once I'm two moves up, I've forgotten everything past the third hold''}. Route information might be given as a rapid overview, as a sequence of local instructions, or only in response to specific questions. As a result, route preview was often fragmented. 

Secondary stakeholders described a parallel challenge from the support side. Sighted guides noted that they often improvised route explanations and that climbers could struggle with cognitive overload. \textit{``You can see them get overloaded,''} one coach explained. 
% \textit{``I describe ten moves, but they can only hold three in their working memory before they touch the wall''} (Guide 1). 
\textit{``I describe ten moves, but they can only keep three in mind when they get on the wall''} (Coach 1).

In summary, these accounts suggest that route preview support should not be treated as a single one-shot description, but as a structured process that helps climbers first form an overview and then access detail as needed.

%\subsubsection{Real-Time Guidance Can Easily Overload Timing and Action}
\subsubsection{Real-time guidance can easily overload timing and action}

Participants generally viewed real-time prompting as useful, but emphasized that its value depended heavily on timing and information density. Guidance that was too long or poorly synchronized could interrupt rather than support action. \textit{``If someone talks for five seconds describing a hold while I'm hanging by one arm, I'm already pumped out. Just tell me `right hand, two o'clock' so I can move,''} explained P1. 

Participants distinguished between helpful and unhelpful prompting less by whether the support was \textit{detailed}, but more by whether it was actionable \textit{in the moment}. Short prompts that fit the temporal demands of climbing were preferred. Secondary stakeholders echoed this, noting that phrasing that looked sufficient in theory could become ambiguous on the wall. For example, phrases such as \textit{``above you''} or \textit{``a little to the right''} were often too underspecified unless paired with distance bands or clock-face references, because climbers could not easily determine how far or in which precise direction to search while under load. Effective prompting was treated as a coordination practice prioritizing low-overload cueing that is brief, repeatable, and tightly aligned with action timing.

%\subsubsection{Uncertainty Must Be Explicitly Managed}
\subsubsection{Uncertainty must be explicitly managed}

A third theme concerned how uncertainty was experienced and handled. Participants described uncertainty as not only losing track of a hold, but also doubting a prompt or lacking a clear recovery path. What seemed especially undesirable was a system or guide that remained overly confident when wrong. \textit{``The worst thing isn't silence; it's a guide or an app acting 100\% sure when it's actually lost. I need to be able to say `wait, let me check' or `say that again' without the system plowing ahead,''} emphasized P4.

Participants' responses suggested that uncertainty handling should be treated as part of support design rather than an exceptional failure case, echoing classic human-automation research showing that poorly calibrated trust and over-reliance on automation can undermine monitoring and recovery when systems behave with unwarranted certainty \cite{Lee2004TrustAutomation,Parasuraman1997HumansAutomation}. Secondary stakeholders agreed, noting that the responsible response to ambiguity was often to simplify, clarify, or explicitly take a more conservative stance. A future system should make uncertainty legible and support a transition to safer, human-assisted modes when needed.

%\subsubsection{Accessibility Support Must Preserve Pace, Enjoyment, and Autonomy}
\subsubsection{Accessibility support must preserve pace, enjoyment, and autonomy}

The final theme concerned what participants considered meaningful support beyond task completion alone. BLV climbers frequently framed good support in terms of maintaining rhythm, preserving exploration, and reducing unnecessary dependence. \textit{``I don't want to be a remote-controlled climber,''} P2 noted. \textit{``Half the fun is solving the puzzle myself. I just want the basic info so I don't have to keep yelling down to ask where the next pink hold is.''}

Participants objected to assistance that felt overly controlling, socially burdensome, or eliminated self-directed problem-solving. They wanted less continuous dependence for ordinary route information and more control over when support was requested. Accessible climbing support should therefore balance usefulness with pace, enjoyment, and user control.  Table~\ref{tab:design_goals_mapping} summarizes how these four findings were translated into design goals for \textit{HoldLight}, linking the formative inquiry to the system design in Section~\ref{sec:system-design}.

%\subsection{Design Goals for HoldLight}

\begin{table*}[h]
\caption{Mapping of Phase A formative findings to the derived design goals for HoldLight.}
\label{tab:design_goals_mapping}
\centering
\small
\tagpdfsetup{table/header-rows={1}}
\begin{tabular}{p{0.44\linewidth} p{0.49\linewidth}}
\toprule
\textbf{Formative study findings} & \textbf{Derived design goals (DGs)} \\
\midrule
\textbf{Finding 1:} Route preview was fragmented, socially dependent, and difficult to retain when overloaded.
& \textbf{DG1: Support layered, on-demand route preview.} Provide an overview-first structure and allow users to request additional detail only when needed, rather than forcing all route information into a single pre-climb explanation. \\
\addlinespace
\textbf{Finding 2:} Real-time guidance was useful but could easily overload timing, balance, and action.
& \textbf{DG2: Minimize overload in real-time next-hold cueing.} Use concise, actionable, and repeatable cues aligned with bodily pacing rather than dense continuous narration. \\
\addlinespace
\textbf{Finding 3:} Uncertainty needed to be explicitly managed rather than hidden.
& \textbf{DG3: Support explicit uncertainty handling and fallback.} Make uncertainty legible and provide repeat, more-detail, pause, and human-help recovery paths. \\
\addlinespace
\textbf{Finding 4:} Accessibility support had to preserve pace, enjoyment, and autonomy rather than merely maximize instruction.
& \textbf{DG4: Preserve climbing pace, enjoyment, autonomy, and bounded human collaboration.} Reduce repeated immediate human re-entry for routine route-information clarification while retaining human safety support and high-uncertainty human judgment. \\
\bottomrule
\end{tabular}
\end{table*}

\section{System Design}
\label{sec:system-design}

\begin{figure*}[t]
  \centering
  \includegraphics[
  width=\textwidth,
  alt={A five-column workflow diagram showing the end-to-end HoldLight route-information journey. From left to right, the stages are Wall Capture, Route Setup and Confirmation, Layered Preview, In-Climb Cueing, and Fallback and Re-entry. Three horizontal rows distinguish user actions, system pipeline, and output or control. Wall capture produces a wall scan; hold detection and color correction support route setup; an ordered route representation drives layered preview; pose tracking, wall alignment, and session state drive cue generation; and confidence checks trigger conservative support or bounded human route-information re-entry. A bottom safety-boundary band indicates that human safety oversight remains available throughout.}
  ]{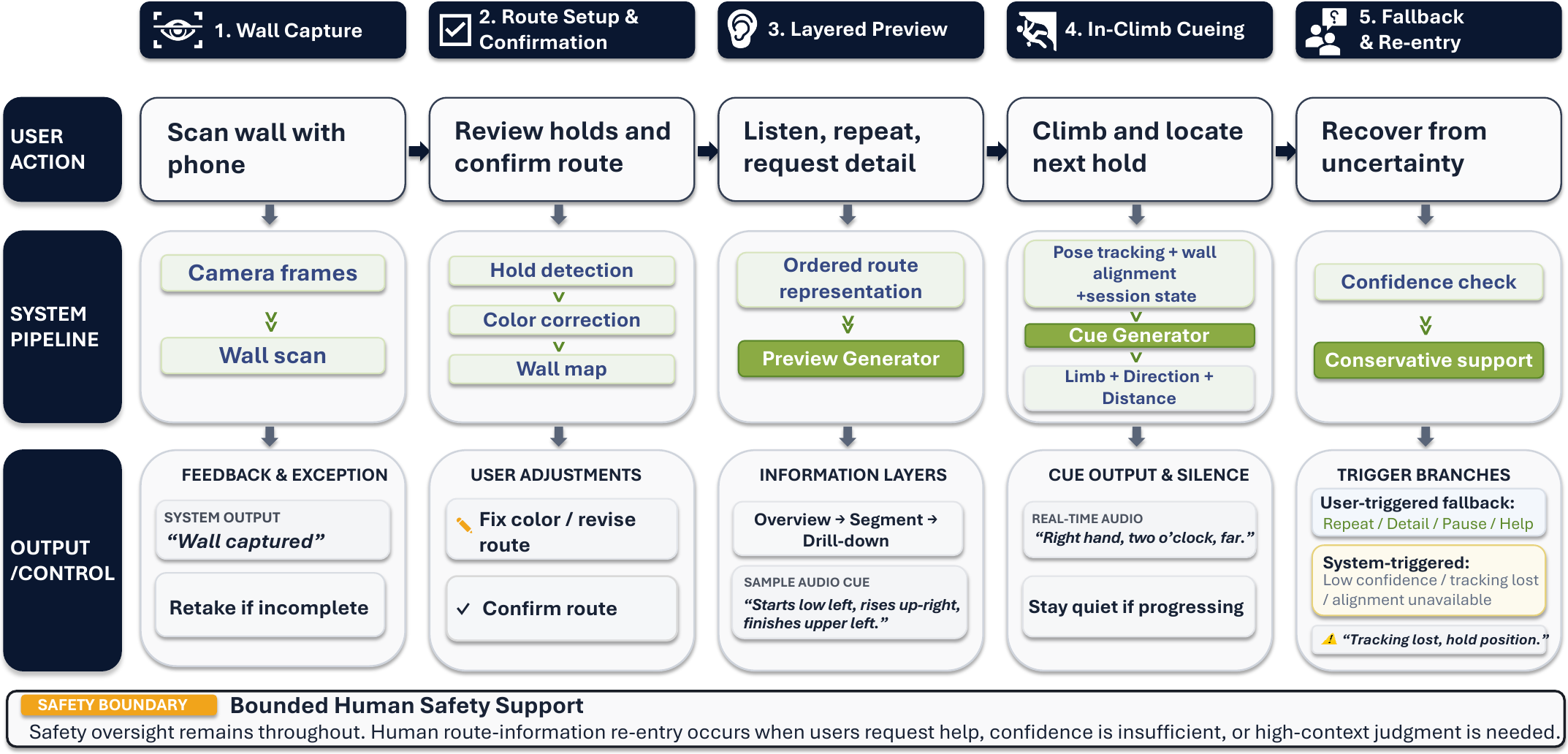}
  \caption{System overview and route-information journey of \textit{HoldLight}. The figure traces how wall capture is converted into an ordered route representation, delivered through layered preview and concise in-climb cues, and shifted into explicit fallback or bounded human route-information re-entry when uncertainty exceeds the system boundary. User actions, system processing, and user-facing outputs or controls are shown at each stage, while human safety oversight remains available throughout.}
  \Description{A five-column workflow diagram showing the end-to-end HoldLight route-information journey. From left to right, the stages are Wall Capture, Route Setup and Confirmation, Layered Preview, In-Climb Cueing, and Fallback and Re-entry. Three horizontal rows distinguish user actions, system pipeline, and output or control. Wall capture produces a wall scan; hold detection and color correction support route setup; an ordered route representation drives layered preview; pose tracking, wall alignment, and session state drive cue generation; and confidence checks trigger conservative support or bounded human route-information re-entry. A bottom safety-boundary band indicates that human safety oversight remains available throughout.}
  \label{fig:system-overview}
\end{figure*}

To address \textbf{RQ2}, we designed \textit{HoldLight} around the four design goals derived from Phase A. Fig.~\ref{fig:system-overview} presents the resulting five-stage route-information workflow: wall capture, route setup and confirmation, layered preview, in-climb cueing, and fallback with bounded human re-entry. The figure distinguishes user actions, system processing, and user-facing outputs or controls, while human safety support remains available throughout.

Before route information is exposed, the system applies lightweight hold-color correction and route revision to improve representation stability.

The workflow is enabled by wall scanning, hold detection, rule-based route planning, browser-side pose tracking, server-assisted wall alignment, spoken cue delivery, and session-state management. Implementation details are provided later in this section and in Appendix~\ref{sec:appendix-runtimearchi}.

\subsection{DG1: Support Layered, On-Demand Route Preview}

\begin{figure}[htbp]
  \centering
  \includegraphics[
  width=\linewidth,
  alt={An illustration showing three levels of user-interface detail on a mobile screen. First, 'Overview-First Preview' shows a simplified path highlighting the start, end, and major turns. Second, 'Segment-Level Preview' breaks the path into three highlighted color blocks. Third, 'On-Demand Drill-Down' zooms into a specific cluster of holds to show a detailed transition move.}
  ]{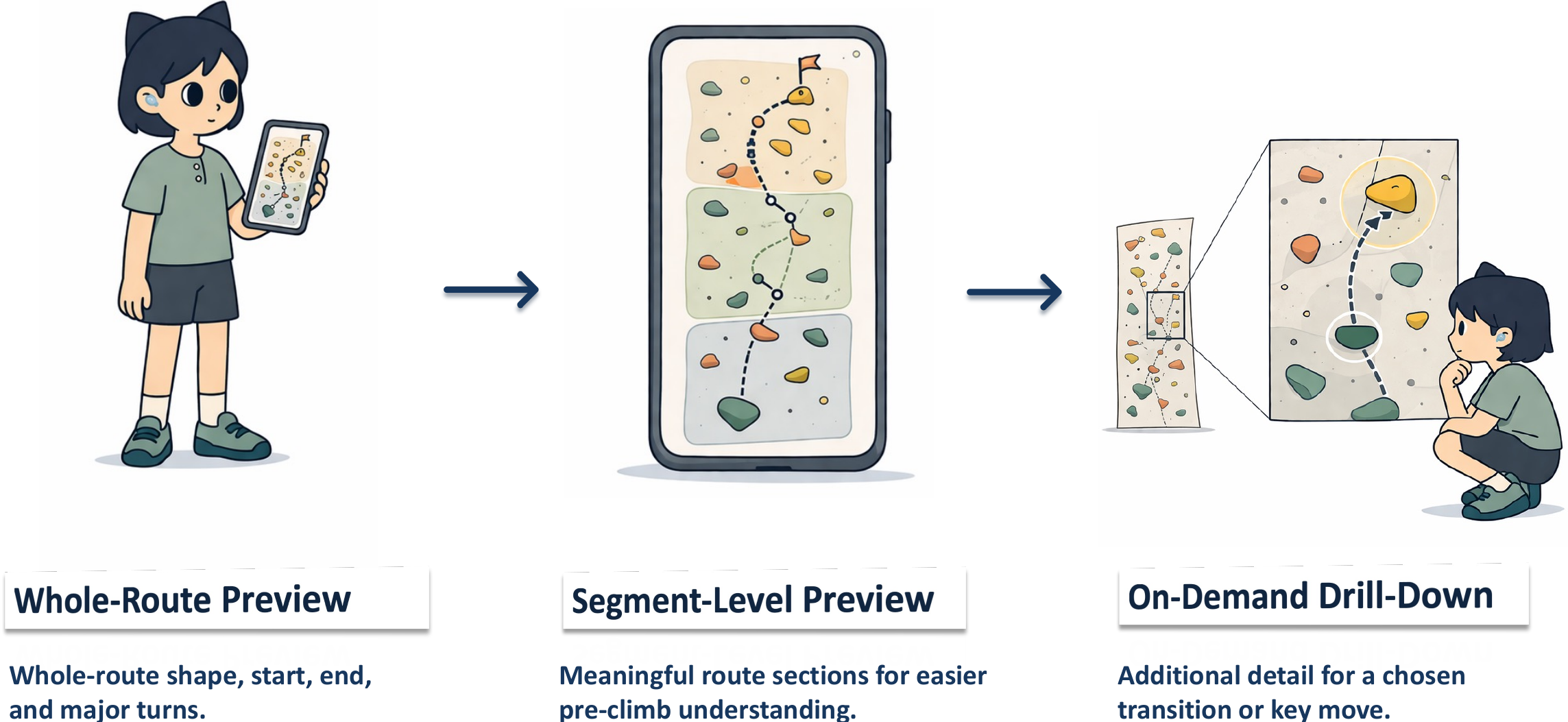}
  \caption{Layered route preview supporting DG1. The interface progresses from a whole-route overview to segment-level preview and optional drill-down for selected transitions.}
  \Description{An illustration showing three levels of user-interface detail on a mobile screen. First, 'Overview-First Preview' shows a simplified path highlighting the start, end, and major turns. Second, 'Segment-Level Preview' breaks the path into three highlighted color blocks. Third, 'On-Demand Drill-Down' zooms into a specific cluster of holds to show a detailed transition move.}
  \label{fig:layered-preview}
\end{figure}

Phase A showed that route understanding before climbing was often fragmented, socially dependent, and difficult to retain when too much information was delivered at once. We therefore designed the preview layer around an overview-first interaction model. Fig.~\ref{fig:layered-preview} illustrates the overview-first preview structure, which lets users inspect segments or transitions in more detail only when needed. The goal was not to exhaustively describe every move before the climb, but to help the user form a usable initial mental representation of the route and then access additional detail only when needed.

The preview is organized into progressively more specific layers. This design was informed by broader BLV accessibility work that emphasizes exploration-oriented and context-sensitive support rather than a single fixed presentation \cite{Kuribayashi2025WanderGuide,Mohanbabu2024ContextAwareImageDescriptions,Chang2024WorldScribe}. At the highest level, the system provides a concise route overview that identifies the intended route, the start and end region, the overall route direction, and the major turning points or salient segments. This first layer is intentionally coarse. It is designed to answer the question, \textit{``What is the general flow of this route?''} rather than \textit{``What should I do at every move?''}

In the second layer, the user can request segment-level preview. Here the route is broken into a small number of meaningful sections, such as lower, middle, and upper segments, or other route-dependent partitions anchored around major transitions. Segment-level preview helps the user understand where the route changes direction, where a key transition occurs, and which parts may require more attention later in the climb.

In the third layer, the user can request more specific information about a chosen segment or key move. This drill-down layer is used only on demand. It allows the system to provide additional detail about a particular transition, a likely difficult section, or the next part of the route that the user wants clarified before climbing begins. By keeping this detail optional and on demand, the system follows a broader accessibility design direction toward context-sensitive and user-adjustable support \cite{Mohanbabu2024ContextAwareImageDescriptions,Chang2024WorldScribe}. This avoids a single maximal description for every situation.

User control is central to this design. The preview does not advance on its own as a long monologue. Instead, the user can step through the route at their own pace, request repetition, or ask for more detail for a specific segment. This interaction model directly addresses the Phase A finding that useful preview is not simply a matter of more information, but of receiving route information in a form that supports both comprehension and later recall.

\subsection{DG2: Minimize Overload in Real-Time Next-Hold Cueing}

Phase A also showed that real-time guidance could be helpful but could easily become disruptive if it was too long, too late, too abstract, or poorly matched to the user's current bodily state. We therefore designed the in-climb guidance layer around short, repeatable, action-oriented cues rather than continuous narration.

The default cue style is intentionally brief. Instead of attempting to verbally describe the entire local scene, the system provides a compact cue directed toward the next actionable target. These cues use a constrained and consistent vocabulary centered on body-relevant spatial reference, such as relative direction, coarse height relation, and limited distance language. Drawing on prior wayfinding and mobility literature on structured messaging and scene description, we designed a restricted cue vocabulary intended to lower interpretive burden during climbing and to provide a more stable support language across routes \cite{vanDerBie2019Sidewalk,Scheuerman2017LearningLanguage,Hoogsteen2022BeyondTheCane}.

Cue timing is equally important \cite{Vincenzi2021InterdependenceInAction,Barbareschi2024SpeechIsSilver}. The system is not designed to speak continuously while the climber is moving. Instead, it issues a cue only when the interaction state indicates that support is needed: for example, when the user requests guidance, when progress inference suggests the current move has ended and the next target is relevant, or when the user appears to have lost track of the next hold. If the user is already actively executing a move or appears to be progressing without difficulty, the system remains quiet rather than competing with ongoing action.

Repetition is treated as a first-class part of the cueing strategy. A brief cue can be replayed when requested, and if the user still needs support, the system can shift from the default short cue to a slightly richer clarification. This progression from brief support to optional elaboration is deliberate. It allows the system to preserve rhythm in the common case while still offering additional support when the short cue alone is insufficient.

By design, the system does not aim to fully narrate the climb. A dense narration strategy might appear more informative in the abstract, but it would conflict with the temporal demands identified in Phase A and discussed in the Related Work. For this reason, the real-time layer prioritizes actionable timing and low interpretive burden over verbally complete instructions, consistent with broader work on time-sensitive guidance and navigation cue design \cite{Barbareschi2024SpeechIsSilver,vanDerBie2019Sidewalk,Gordon2023HearingWayForward}. Informed by common sight-guiding practice in BLV climbing and by structured directional cueing strategies identified through formative interviews, the cue format prioritized direction first, then coarse distance, and only brief additional information when needed. Example cue forms included phrases such as \textit{``right hand, two o'clock, medium,''} \textit{``left foot, eleven o'clock, close,''} and short confirmations such as \textit{``yes, that's it''} or \textit{``on target, steady.''}

Table~\ref{tab:route_delivery_examples} extends these examples across the broader route-information workflow, showing how internal route representations were translated into user-facing outputs during preview, in-climb cueing, clarification, and fallback.

\begin{table}[t]
\caption{Examples of route-information delivery across \textit{HoldLight}'s interaction layers.}
\label{tab:route_delivery_examples}
\centering
\footnotesize
\setlength{\tabcolsep}{3pt}
\renewcommand{\arraystretch}{1.05}
\tagpdfsetup{table/header-rows={1}}
\begin{tabular}{p{0.15\linewidth} p{0.38\linewidth} p{0.39\linewidth}}
\toprule
\textbf{Interaction layer} &
\textbf{System representation} &
\textbf{Example user-facing output} \\
\midrule

Overview preview &
Start, finish, overall direction, and major transition &
\textit{``The route starts low on the left side of the wall, trends up and right through the middle, and finishes near the upper center.''} \\

Segment preview &
Hold distribution and direction within the lower, middle, and upper sections &
\textit{``Lower segment: the handholds trend up and right, with a supporting foothold on the left.''} \\

Default next-hold cue &
Limb, clock-face direction, and body-relative distance band &
\textit{``Right hand, two o'clock, far.''} \\

More detail &
Limb, hold color or type, and additional spatial information &
\textit{``Right hand to the blue jug at two o'clock, about one arm's reach away.''} \\

Fallback &
Low confidence or user-requested recovery &
\textit{``Tracking is uncertain. Pause if safe, or request human help.''} \\

\bottomrule
\end{tabular}
\end{table}

\subsection{DG3: Support Explicit Uncertainty Handling and Fallback}

\begin{figure}[htbp]
  \centering
  \includegraphics[
  width=\linewidth,
  alt={A flowchart detailing the system's real-time interaction states. When progressing normally, the system is quiet. When a cue is needed, a brief next-hold cue is provided. If the climber needs more, they can request repetition. If the system faces uncertainty, it provides conservative support rather than guessing, and human help always remains available as a final fallback.}
  ]{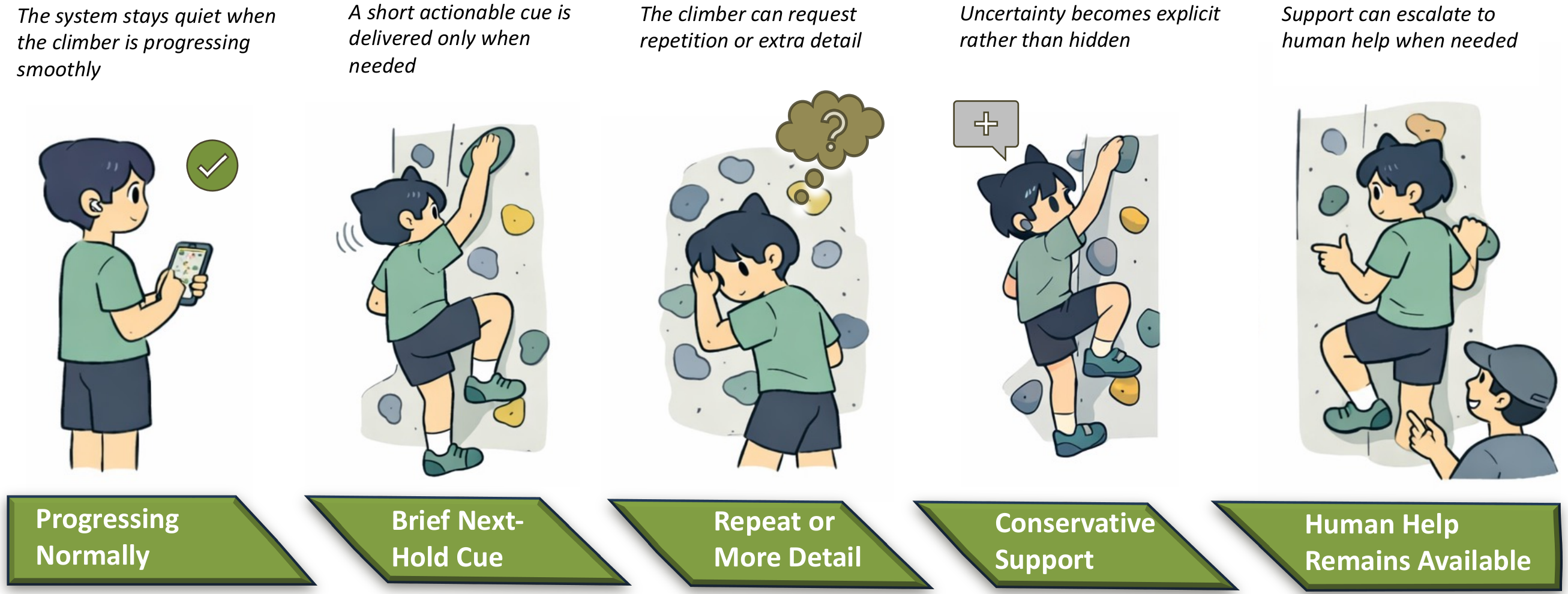}
  \caption{HoldLight's pacing and fallback logic (addressing DG2 and DG3). The system remains quiet during smooth progression, provides short actionable cues when needed, and explicitly escalates to conservative support or human help when uncertainty arises.}
  \Description{A flowchart detailing the system's real-time interaction states. When progressing normally, the system is quiet. When a cue is needed, a brief next-hold cue is provided. If the climber needs more, they can request repetition. If the system faces uncertainty, it provides conservative support rather than guessing, and human help always remains available as a final fallback.}
  \label{fig:fallback-logic}
\end{figure}

Phase A showed that uncertainty should be made legible rather than hidden as an implementation failure. Fig.~\ref{fig:fallback-logic} summarizes the resulting interaction policy, linking concise cueing to explicit recovery when uncertainty arises. In climbing, uncertainty is frequent and consequential: the user may be unsure whether a cue was heard correctly, whether it matches the intended hold, whether the system has lost track of progress, or whether continuing independently remains efficient or comfortable. These conditions motivated the design of fallback as an explicit accessibility layer.

The system supports both user-triggered and system-triggered fallback. On the user side, the interaction provides a small set of explicit recovery actions: repeating the last cue, requesting more detail, pausing momentarily, and requesting human help. These actions are intentionally limited and legible. The goal is not to provide an open-ended conversational interface during climbing, but to make the main recovery paths immediately available when the user becomes uncertain.

On the system side, fallback is also triggered when confidence in route-specific guidance becomes insufficient. This may occur when route tracking becomes unreliable, when progress inference no longer matches recent user movement, when the system cannot identify the user's state with enough confidence to give a target-specific cue responsibly, or when repeated mismatch suggests that continuing with the same cueing strategy would likely increase confusion. In these cases, the system does not continue issuing overconfident next-hold prompts. Instead, it shifts into a conservative support mode.

In conservative mode, the system makes its uncertainty explicit and changes its behavior accordingly. It may ask the user to pause, offer a repeat of the last stable information, offer a more general orientation cue, or recommend human assistance. This is a deliberate design choice. Prior work argues that uncertainty should be made legible to support users' reliability judgments and recovery decisions \cite{Chen2025SurfacingVariations}. In our design, the system therefore treats uncertainty as information that should be surfaced rather than concealed.

This explicit fallback logic is important not only for safety and trust, but also for autonomy \cite{Kamikubo2025BeyondOmakase,Ranganeni2023ExploringLevelsOfControl,Alharbi2024MisfittingWithAI}. By giving the user clear ways to recover without surrendering control immediately, the system supports a graded interaction in which uncertainty can be managed rather than simply endured.

\subsection{DG4: Preserve Climbing Pace, Enjoyment, and Autonomy}

Finally, Phase A showed that helpful support in climbing is not defined solely by whether it helps the user reach the top. Participants also cared about rhythm, naturalness, enjoyment, and the ability to keep the climb recognizably their own. We therefore designed the overall interaction so that the system assists without dominating.

The first principle is selective intervention. The system does not attempt to direct every micro-decision. Route preview provides structure before the climb, and real-time guidance is available during the climb, but the system is intentionally quiet unless support is needed or requested, aligning this behavior with a mixed-initiative interaction stance in which automated services remain available while users retain control over when and how assistance is invoked \cite{Horvitz1999MixedInitiative}. This preserves room for self-directed exploration and reduces the risk that the interaction becomes an externally imposed script rather than a support tool.

The second principle is user control over support depth. Informed by broader BLV accessibility systems that prioritize context-sensitive and customizable assistance \cite{Kuribayashi2025WanderGuide,Chang2024WorldScribe,Mohanbabu2024ContextAwareImageDescriptions}, \textit{HoldLight} allows users to request repetition or more detail across preview, cueing, and fallback, without imposing maximal detail by default. This supports autonomy by allowing users to determine when the current level of assistance is sufficient and when escalation is necessary.

The third principle is bounded human collaboration. Human safety support remains present, but the system is designed to handle some routine route-information clarification before immediate human re-entry is needed. This preserves a clear boundary for when human help should re-enter the interaction: not because the system replaces human judgment, but because routine prompts, repetition, and low-level clarification can sometimes be handled through the workflow while safety-critical or highly situational moments remain human-led.

\subsection{Implementation}

\begin{figure*}[t]
  \centering
  \includegraphics[
  width=\textwidth,
  alt={A grid of eight illustrated cards showing the system setup. The top row shows handheld wall capture, route scanning, route choosing on a phone, and a stationary phone on a tripod facing the climbing wall. The bottom row shows browser-side pose tracking of the climber's skeleton, server-assisted route processing of holds, a climber wearing a single-ear audio device, and a human spotter providing bounded safety support below.}
  ]{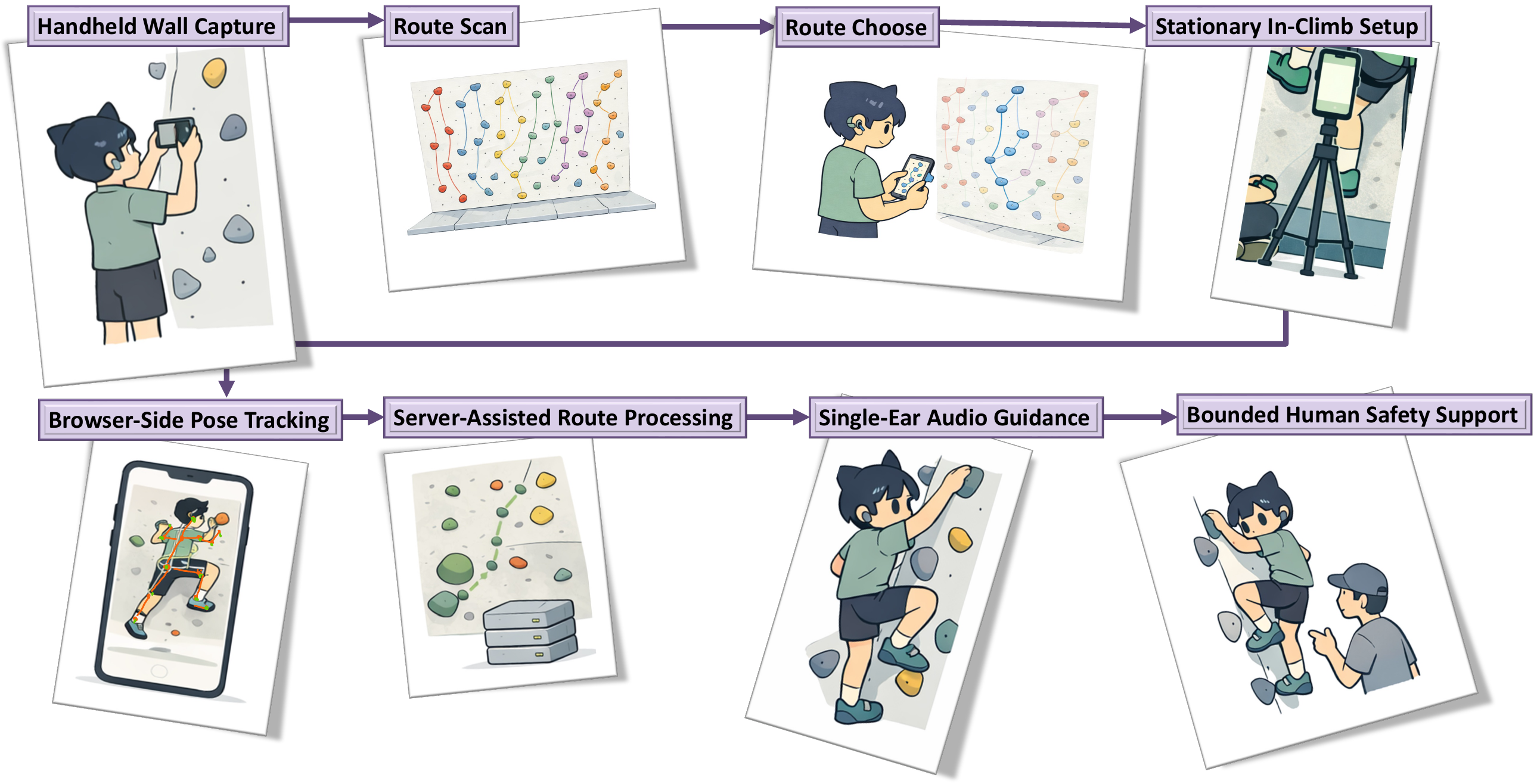}
  \caption{Apparatus and implementation workflow. In the deployed study setup, browser-side pose tracking supports local state estimation, server-assisted processing handles route alignment, and bounded human safety support remains part of the overall workflow.}
  \Description{A grid of eight illustrated cards showing the system setup. The top row shows handheld wall capture, route scanning, route choosing on a phone, and a stationary phone on a tripod facing the climbing wall. The bottom row shows browser-side pose tracking of the climber's skeleton, server-assisted route processing of holds, a climber wearing a single-ear audio device, and a human spotter providing bounded safety support below.}
  \label{fig:implementation}
\end{figure*}

As shown in Fig.~\ref{fig:implementation}, the deployed setup relies on a browser-client and server-assisted workflow. The implementation was designed to support the interaction workflow described above, with evaluation focused on workflow usability rather than benchmark-style recognition performance. \textit{HoldLight} was implemented as a browser-based mobile web application and, in our study setup, was accessed on an iPhone 15. The phone's built-in camera was used for both pre-climb route preview and in-climb guidance. During preview, the phone could be hand-held to capture the wall; during climbing, the phone remained stationary, either placed at ground level or fixed on a tripod. The same camera input pipeline was used across preview and in-climb support.

Audio guidance was delivered through single-ear wearable audio output, using either a single-ear Bluetooth earpiece or a single-sided bone-conduction device. Prior work notes that bone-conduction headphones leave the ears unobstructed and preserve normal auditory acuity \cite{Schoop2018HindSight,Kayukawa2020GuidingBlindPedestrians,Sato2019NavCog3InTheWild,Kamikubo2020SupportStrategiesRemoteGuides,Kameswaran2020UnderstandingInSituUse}. We therefore used single-ear and open-ear audio delivery to preserve access to environmental sounds and safety-related verbal cues in the climbing environment.

The full processing path used a browser-plus-server architecture. A lower-level view of the deployed runtime architecture is provided in Appendix~\ref{sec:appendix-runtimearchi}, which details the browser-centric implementation stack, the split between browser-local capabilities and server-assisted processing, and the main wall-scan and live-guidance pipelines. In our implementation, browser-side components handled pose tracking, spoken cue delivery, and the user-facing interaction logic, while wall-scan recognition, route extraction, and live wall-alignment requests were server-assisted. In practice, the browser captured images or image frames and sent vision requests to backend services, while scan, session, and interaction records were stored server-side.

For pose-based state estimation, the front end used the MediaPipe Pose library, building on the MediaPipe perception-pipeline framework and the BlazePose / BlazePose GHUM family, for browser-side body tracking \cite{Lugaresi2019MediaPipe,Grishchenko2022BlazePoseGHUM}. In our implementation, browser-side pose tracking was used to estimate coarse body state and action progression, not to directly recognize target holds from the live camera view. When pose quality degraded---for example, because of body self-occlusion, transient multi-person interference, or insufficient visible joints/limbs---the live tracker did not continue issuing equally confident movement-state updates. Instead, the interaction could downgrade into more conservative states such as holding, reacquiring, or searching until sufficient pose evidence returned.

Live target-hold visibility during climbing was handled through server-assisted wall alignment rather than through browser-side pose inference. At a fixed refresh interval, the browser captured JPEG frames and sent them to the backend, where the system attempted to align the current camera view to the previously scanned wall map using feature matching followed by homography estimation with RANSAC \cite{FischlerBolles1981RANSAC}. If the alignment evidence was insufficient---for example because too few matches were available, the inlier ratio was too low, or reprojection error exceeded the acceptance threshold---the backend returned a partial or unavailable alignment state rather than treating the wall registration as successful. In those cases, the front end could suppress target-specific cueing and rely on repeat, retake, or fallback behavior instead of continuing with overconfident guidance.

Before interaction, the system constructed a route representation used for preview and later progress inference. The ordered route used for live guidance was generated by a custom browser-side rule-based semantic route planner rather than by a learned planner or an external heuristic library. In simplified terms, the planner first formed same-color route candidates, then applied reachability constraints that varied by route difficulty and local progression, including bounded lateral reach, vertical gain/drop, and start-traverse checks. It then performed a bounded search over candidate handhold sequences, inserted support holds when necessary, and scored candidate paths using start/finish semantic checks together with reachability and local stability criteria. The resulting ordered route was therefore not a purely visual ordering of holds, but a rule-based sequence intended to remain compatible with climbable progression and later target-specific cueing.

The planner produced a hand-primary ordered sequence over same-color candidate holds. When a move required lower-body support or when hand-only progression would violate local reachability or stability constraints, the planner inserted selected support holds as foot targets. The cue generator then assigned left/right hand or foot labels using target side, previous limb use, clock-face direction, and coarse distance bands. We did not evaluate planner optimality against expert-selected route sequences; our claims concern the usability and recoverability of the route-information workflow rather than expert-level route planning.

During climbing, browser-side pose tracking and server-assisted wall alignment jointly informed whether the system should remain silent, issue a brief next-hold cue, or shift into a more conservative fallback state. In our study implementation, pose inference was scheduled on the browser side at approximately sub-second frequency, whereas wall-alignment refresh operated at a slower multi-second interval. We did not maintain end-to-end latency logs for the full camera-to-server-to-audio path and therefore do not claim a formal latency benchmark. Instead, timing-related observations in Phase C should be interpreted as workflow-level evidence about when the deployed system remained actionable and when it degraded into delayed cues, retakes, or fallback.

The user-facing interaction exposed a compact set of accessible actions across preview, cueing, and fallback, including repetition, more detail, pause, and human help. In the deployed study, these actions were invoked through the short commands \textit{repeat}, \textit{detail}, \textit{pause}, and \textit{help}. These controls were intentionally limited so that the interaction could preserve pacing and user control rather than encourage open-ended dialogue during climbing.

Because the user-facing value of \textit{HoldLight} depended on the stability of route representation and cue timing, the Phase C evaluation also tracked a small set of technical and interactional reliability events. These included route-correction or route-revision events prior to interaction, delayed or ambiguous target-specific cues, user-triggered and system-triggered fallback transitions, and cases where human information support re-entered the interaction following degraded system confidence. These events were not treated as a recognition benchmark, but as contextual evidence for interpreting where the workflow remained effective and where it degraded.

\section{Phase B: Pilot Refinement}

\subsection{Procedure}

We recruited 4 BLV climbers (two blind, two low-vision), using a small targeted pilot sample because the purpose of this phase was iterative refinement rather than outcome comparison. Each pilot session included three short tasks: (1) a pre-climb preview task, where participants listened to an overview and then described what they understood about the route; (2) a short guided-climb task, where participants used next-hold cues under bodily load; and (3) a debrief comparing cue length, timing, and fallback clarity. We used rapid post-task debriefs and structured observation to identify moments of hesitation, repeated clarification, delayed action, and fallback use. These observations were used to decide which interaction variants were sufficiently stable for inclusion in Phase C.

\subsection{Pilot Findings}

The pilot revealed three recurring practical issues. First, monolithic pre-climb route explanations were quickly forgotten once the user was under bodily load. Second, unconstrained cue phrasing caused dangerous cognitive delays during climbing, consistent with prior work suggesting that tasks demanding substantial cognitive processing capacity can interfere with ongoing performance when working-memory resources are limited \cite{Sweller1988CognitiveLoad,BaddeleyHitch1974WorkingMemory}. For example, a verbose prompt like \textit{``Move your right hand to the blue jug at 2 o'clock, about 30 centimeters away''} required too much interpretation time, often causing the climber to lose balance while processing the instruction. Third, when the prototype lost tracking confidence without informing the user, climbers continued searching for nonexistent holds based on outdated cues, accumulating confusion rather than recovering. In early pre-formal testing, server-side resource limitations also produced occasional end-to-end delays that prevented timely next-hold cue delivery. In one such case, the climber discontinued the attempt because the delayed cueing was no longer actionable under bodily load. We therefore upgraded the server configuration before Phase C and did not include these pre-formal trials in the formal evaluation.

\subsection{Design Revisions}

Based on these pilot findings, we finalized four interaction revisions before Phase C: a layered preview hierarchy (Overview $\rightarrow$ Segment $\rightarrow$ Drill-down), a stripped-down ``Direction-First, Distance-Second'' cue format, silence during active move execution, and explicit fallback actions and states. For example, the overly verbose cue mentioned above was reduced to \textit{``Right hand, two o'clock, far.''} Likewise, instead of passively timing out when confidence dropped, the system now exposed explicit fallback through user actions such as \textit{``Repeat''} and \textit{``More Detail''} and system prompts such as \textit{``Tracking lost, hold position.''} Together, these revisions stabilized the interaction for formal comparison in Phase C without changing the study’s bounded focus on route-information support.

Because the pilot included only two blind and two low-vision participants, we treated any vision-status differences as design observations rather than subgroup findings. Across this small pilot, the most consistent pattern was not a stable blind/low-vision split, but a shared preference for shorter cues and explicit recovery options during physically loaded moments.

\section{Phase C: Controlled Initial Evaluation}

To address \textbf{RQ3a} and \textbf{RQ3b}, we conducted a controlled initial evaluation of the system in indoor climbing tasks using a within-subjects design. The goal of this phase was to compare two bounded information-support workflows under controlled indoor climbing conditions, not to determine whether \textit{HoldLight} should replace human guides or human safety personnel. Accordingly, we treat this study as an initial mixed-method evaluation of accessibility support under controlled conditions rather than as a definitive effectiveness trial. \textbf{RQ3a} focuses on structured route understanding and next-hold recovery, whereas \textbf{RQ3b} focuses on perceived autonomy and immediate human re-entry for routine route-information clarification.

\subsection{Study Design and Conditions}

We employed a two-condition within-subjects design. Each participant completed both a controlled on-demand human-support condition and the \textit{HoldLight} condition. Condition order was counterbalanced, and the two conditions used difficulty-matched routes rather than the exact same route to reduce simple memorization and transfer. Each participant completed three formal routes per condition, in addition to one familiarization route used only for training and not included in formal analysis.

\subsubsection{Condition A: Controlled On-Demand Human-Support Condition}

Condition A was designed as a controlled on-demand human-support condition derived from current route-information support practices. It was not intended to model the full flexibility of unconstrained expert coaching. Instead, it provided a reproducible comparison for examining how two bounded route-information workflows supported route understanding and next-hold recovery. For concision, we refer to this controlled on-demand human-support condition as the baseline in the results and tables.

The same trained guide delivered this controlled baseline condition throughout the study.
 Prior to data collection, the guide was calibrated to a predefined prompt policy: delivering a pre-climb explanation covering the start, end, overall direction, and major transitions and providing in-climb guidance only upon explicit participant request or following a visually confirmed breakdown (e.g., a major pause lasting $\geq 6$ seconds under the study coding rules).

To monitor consistency, the guide wore a lapel microphone. A secondary researcher served as a moderator, using a fidelity checklist to score whether the guide’s interventions remained within the predefined boundaries. Post-session reviews confirmed that the guide adhered to the intended prompt policy in 34 of the 36 baseline trials (94.4\%). The two deviations involved the guide preemptively offering a directional cue before the 6-second pause threshold was reached. These two trials were retained in the analysis because the deviations were minor and did not alter the shared safety boundary or the overall condition structure. This compliance rate supports interpreting the baseline as a consistent controlled comparison.

\subsubsection{Condition B: HoldLight}

Condition B used \textit{HoldLight}, including layered pre-climb route preview, concise real-time next-hold guidance, and explicit fallback and uncertainty handling. The system was intended to handle a bounded portion of routine route-information clarification while keeping human safety support and high-uncertainty human judgment available throughout.

\subsection{Participants and Ethics}

Phase C participant characteristics are also summarized in Table~\ref{tab:participant_characteristics}. We recruited 12 blind and low-vision participants for the controlled initial evaluation. Eligibility criteria required participants to be aged 18 or older, to self-identify as blind or low-vision, and to have at least two prior indoor climbing experiences. We included participants with diverse visual conditions and climbing experience levels because the study aimed to evaluate accessibility support for route preview and in-climb guidance across a realistic range of BLV climbing backgrounds, rather than only among expert climbers.

As noted in Phase A, all participant-facing phases were covered by institutional ethics approval. Prior to Phase C participation, all participants completed informed consent using accessible communication procedures. Participants were informed that they could pause, request clarification, skip questions, or stop the study at any time without penalty.

Because climbing is a physically and safety-sensitive activity, all formal tasks were conducted in a controlled indoor climbing environment under continuous professional or trained human safety oversight.
% Safety, moderation, and recording roles were separated as far as practical within the study team. Human safety support remained available throughout both conditions and could override the information-support protocol whenever necessary.

The protocol specified that a session would be paused or terminated if any of the following occurred: (1) participant request to stop; (2) visible distress, pain, or loss of confidence judged incompatible with continued participation; (3) unsafe loss of control or unsafe body position identified by safety staff; (4) inability to continue safely after repeated failed recovery attempts; or (5) equipment or communication failure that compromised safe operation. None of the Phase C sessions met these predefined pause-or-termination criteria.

\subsection{Setting, Routes, and Apparatus}

The study was conducted in a commercial indoor climbing gym during standard operating hours. Formal tasks used a predefined set of indoor routes selected to support controlled comparison.

We used six top-rope routes graded between 5.8 and 5.10c on the Yosemite Decimal System \cite{murdock2004jtnpclimbing}, a common climbing difficulty scale in which higher grades within this range indicate more technically demanding routes, paired into three difficulty-matched sets. Routes were paired primarily by climbing grade and approximate layout complexity where feasible. We attempted to avoid extreme same-color density or obvious route intersections, although some local dense clusters and adjacent-route ambiguity inevitably remained, as later reflected in our system boundary observations. Before formal data collection, the research team defined key route segments for each formal route to support consistent scoring of route-understanding and progression-related measures. These key segments were used only as analytic anchors; they did not change the climbing task itself.

\begin{figure*}[t]
  \centering
  \includegraphics[
  width=\textwidth,
  alt={Three photos from the controlled initial evaluation in an indoor climbing gym. In panel (a), a climber is on the wall while a tripod-mounted phone faces the route from the ground and a guide monitors the deployed interface. In panel (b), several close and mid-range views show participants climbing formal top-rope routes under bodily load. In panel (c), a close-up view shows the phone screen displaying the core HoldLight interface used during the deployed workflow.}
  ]{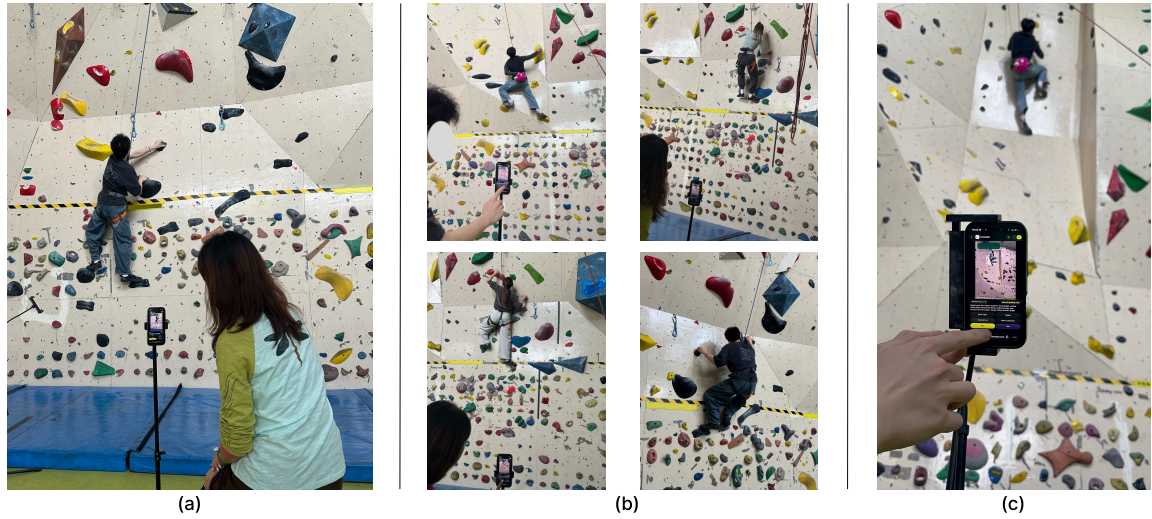}
  \caption{Representative moments from the controlled initial evaluation in the commercial indoor climbing gym. (a) shows the deployed Phase~C setup during a formal trial, including the tripod-mounted phone, the guide monitoring the interface from the ground, and the climber on the wall. (b) shows representative climbing moments from formal top-rope trials, illustrating the bodily-load context in which route understanding, next-hold localization, and uncertainty recovery were assessed. (c) shows a close-up of the phone interface used in the deployed workflow. }
  \Description{Three photos from the controlled initial evaluation in an indoor climbing gym. In panel (a), a climber is on the wall while a tripod-mounted phone faces the route from the ground and a guide monitors the deployed interface. In panel (b), several close and mid-range views show participants climbing formal top-rope routes under bodily load. In panel (c), a close-up view shows the phone screen displaying the core HoldLight interface used during the deployed workflow. }
  \label{fig:phasec_context}
\end{figure*}

The \textit{HoldLight} condition used the complete climbing guidance setup described in Section~\ref{sec:system-design}, including the preview interface, real-time audio guidance, and user-invoked fallback actions. During formal \textit{HoldLight} trials, the phone was tripod-mounted in front of the wall, and the guide monitored the deployed interface from the ground while the climber ascended. The controlled on-demand human-support condition used a human-guidance protocol derived from current support practices but intentionally constrained for reproducible comparison. In both conditions, participants could request repetition, clarification, pausing, or help according to the rules explained during familiarization. Fig.~\ref{fig:phasec_context} shows the deployed Phase~C setup and representative moments from the formal climbing tasks, grounding the controlled evaluation in the real indoor wall context, bodily-load conditions, and phone-based workflow under which route-information support was assessed.

\subsection{Procedure}

Each session followed the same overall sequence: arrival and consent, background confirmation and safety briefing, familiarization, the first assigned condition in the counterbalanced order, rest, the second condition, and a final comparative interview.

\paragraph{Arrival and briefing.}
At the start of each session, we confirmed consent, participant background, and safety procedures.

\paragraph{Familiarization.}
Before formal tasks began, participants completed a familiarization route and a short training interaction appropriate to the upcoming condition. Familiarization covered not only the interaction commands, but also the distinction between ordinary information support and safety override.

\paragraph{Per-condition procedure.}
Each condition included two stages: (1) a pre-climb route-understanding stage and (2) a formal climbing task. In the pre-climb stage, participants received route information using the condition-specific support structure and then completed the route-understanding task. In the climbing stage, participants attempted the formal route while receiving condition-specific support according to the relevant protocol. After each condition, participants completed an immediate post-condition questionnaire.

\paragraph{Rest between conditions.}
A minimum rest period of 10 minutes was provided between conditions. Participants then completed the remaining condition in the assigned counterbalanced order. The second condition began only after both participant self-report and researcher confirmation indicated adequate recovery.

\paragraph{Final comparative interview.}
After both conditions were completed, we conducted a semi-structured comparative interview. The interview explicitly asked which aspects of support felt less socially burdensome and which aspects still clearly required human judgment.

\subsection{Measures}

Measures covered structured route access, uncertainty recovery during climbing, immediate human re-entry for routine route-information clarification, and subjective responses concerning autonomy, trust, safety boundary, and acceptability.

\subsubsection{Structured Route Access}

To address the route-understanding portion of \textbf{RQ3a}, participants completed a brief route-understanding task after receiving condition-specific route information and before beginning the formal climb. The task assessed whether participants could form a workable picture of the route, including its start, end, overall direction, and salient transitions or segments. The item-level 4-point scoring rubric used to evaluate these structured features is provided in Table~\ref{tab:route_understanding_rubric} in Appendix~\ref{sec:appendix-studyinstruments}.

\subsubsection{Uncertainty Recovery During Climbing}

To address the next-hold recovery portion of \textbf{RQ3a}, we recorded next-hold-related breakdown events, major pauses, clarification-dependent interruptions, and cases in which participants lost track of the next target hold. In this paper, next-hold recovery was operationalized through these uncertainty-related breakdown and repair measures rather than through recognition accuracy alone. Where feasible, we also summarized the number and duration of uncertainty breakdown episodes and the time required to recover after onset.

\subsubsection{Immediate Human Re-Entry for Routine Route-Information Clarification}

To address the reliance-on-human-support portion of \textbf{RQ3b}, we recorded explicit help requests, extra human information-support interventions, and cases in which the interaction shifted from system-led support back to human information support. Because one intended benefit of \textit{HoldLight} was to reduce immediate human re-entry for routine route-information clarification without removing human safety support, safety-only interventions were distinguished from information-support interventions wherever possible. We do not interpret this measure alone as evidence of reduced overall support need, because repetition and clarification that would be directed to a human in the controlled on-demand condition could be handled through the system in the \textit{HoldLight} condition.

\subsubsection{Perceived Autonomy, Trust, Safety Boundary, and Acceptability}

To address the perceived-autonomy portion of \textbf{RQ3b}, participants completed a short post-condition questionnaire after each condition using five-point agreement items in a Likert-style format \cite{Likert1932Attitudes}. Questionnaire items were analyzed as pre-specified exploratory grouped item scores rather than validated psychometric constructs. These grouped scores covered in-climb cue usefulness and uncertainty recovery, and trust, safety boundary, and acceptability, plus one single-item autonomy outcome. Items related to perceived route access were used mainly as descriptive and interpretive support alongside the rubric-based route-understanding task, rather than as a separate primary questionnaire construct in the main inferential results. Because the questionnaire did not include a direct multi-item Likert cluster for social dependence, reduced reliance on ordinary human information support was interpreted primarily from behavioral and qualitative measures. Full questionnaire items are provided in Table~\ref{tab:questionnaire_items} in Appendix~\ref{sec:appendix-questionnaire}.

\subsection{Observation and Coding}

Researchers completed a structured observation form during each participant--condition trial. Observation focused on route-understanding behavior, pauses and clarification, uncertainty and recovery, prompt quality, extra human intervention, and notable breakdown or repair moments.

Because observation and post-hoc interpretation could be influenced by condition awareness, we used structured observation forms, pre-specified operational definitions, condition-neutral coding rules, and video-based verification of ambiguous cases where available. Exploratory touch was not automatically coded as an error, and prompt-quality issues were coded in a condition-neutral manner. To maintain analytical clarity, we distinguished between system-logged or protocol-recorded events (e.g., manual route revisions prior to climbing, user-triggered fallback actions) and observation-derived events (e.g., major uncertainty pauses, clarification requests, and human information re-entries). Core event definitions are summarized in Table~\ref{tab:event_definitions} in Appendix~\ref{sec:appendix-scoring}. Observation was used primarily as contextual and explanatory evidence to interpret these counts rather than as stand-alone proof of causal superiority.

\subsection{Data Preparation and Analysis}

Quantitative data were aggregated at the participant level to avoid route-level pseudo-replication. For route-level ordinal and count outcomes, we used participant-level median aggregates across the three formal routes within each condition before inferential comparison. Questionnaire outcomes were analyzed at the participant--condition level because each condition was followed by a single post-condition questionnaire. Given the modest sample size and the use of bounded or ordinal measures for several outcomes, we used paired within-subject comparisons appropriate to the scale and distribution of each measure. For small-sample paired comparisons, we used Wilcoxon signed-rank tests and report exact $p$-values together with the test statistic ($W$), consistent with guidance on nonparametric analysis in HCI for within-subjects data and bounded or ordinal response measures \cite{WobbrockKay2016NonparametricHCI}. We distinguish statistically supported findings from descriptive trends when inferential support was limited or when a result was included primarily for contextual interpretation.

 Qualitative interview and observation materials were used to contextualize how and why the two workflows differed, especially in relation to the outcomes targeted by \textbf{RQ3a} (route understanding and uncertainty recovery) and \textbf{RQ3b} (perceived autonomy and immediate human re-entry for routine route-information clarification). Reliability and fallback events were summarized descriptively to contextualize where the workflow remained effective and where it degraded.

\subsection{Results}

\begin{table*}[h]
\caption{Summary of quantitative outcomes comparing the controlled on-demand human-support condition and \textit{HoldLight} conditions ($N=12$). Values are reported as Median [IQR]. Given the small sample size and bounded or ordinal measures, paired comparisons used Wilcoxon signed-rank tests.}
\label{tab:results_summary}
\centering
\small
\setlength{\tabcolsep}{4pt}
\tagpdfsetup{table/header-rows={1}}
\begin{tabular}{p{0.40\linewidth} p{0.15\linewidth} p{0.15\linewidth} p{0.12\linewidth} p{0.08\linewidth}}
\toprule
\textbf{Measure} & \textbf{Baseline} & \textbf{HoldLight} & \textbf{Statistic} & \textbf{$p$} \\
\midrule
\makebox[0pt][l]{\textit{Structured route access}}
& {} & {} & {} & {} \\
Route-understanding score (0--4) & 2.0 [1.5, 2.5] & 3.5 [3.0, 4.0] & $W = 74$ & \textbf{.005} \\
\addlinespace
\makebox[0pt][l]{\textit{Uncertainty recovery and route continuity}}
& {} & {} & {} & {} \\
Uncertainty breakdown episodes (count) & 4.0 [3.0, 5.5] & 2.0 [1.0, 2.5] & $W = 68$ & \textbf{.008} \\
Clarification-dependent interruptions (count) & 3.0 [2.0, 4.0] & 1.0 [0.0, 2.0] & $W = 70$ & \textbf{.005} \\
Total task time (min:sec) & 4:15 [3:45, 5:10] & 4:30 [3:50, 5:15] & $W = 48$ & .420 \\
\addlinespace
\makebox[0pt][l]{\textit{Immediate human re-entry for routine route-information clarification}}
& {} & {} & {} & {} \\
Explicit requests for human route-information help (count) & 3.5 [2.5, 4.5] & 1.0 [0.0, 1.5] & $W = 72$ & \textbf{.003} \\
Safety-only interventions (count) & 0.0 [0.0, 1.0] & 0.0 [0.0, 0.5] & $W = 37$ & .850 \\
\addlinespace
\makebox[0pt][l]{\textit{Subjective ratings}}
& {} & {} & {} & {} \\
Perceived autonomy (single-item outcome, Q11) & 3.0 [3.0, 4.0] & 5.0 [4.0, 5.0] & $W = 76$ & \textbf{.002} \\
In-climb cue usefulness and uncertainty recovery (grouped item score) & 3.8 [3.2, 4.0] & 4.3 [4.0, 4.8] & $W = 62$ & \textbf{.036} \\
Trust, safety, and acceptability (grouped item score) & 4.2 [3.5, 4.8] & 4.5 [4.0, 5.0] & $W = 45$ & .310 \\
\bottomrule
\end{tabular}

\vspace{4pt}
\raggedright
\footnotesize
\textit{Note.} Route-level count events were integer-coded. For each participant and condition, route-level counts were first aggregated across the three formal routes using participant-level medians. Reported medians and IQRs were then calculated over the 12 participant-level aggregates; decimal values therefore reflect group-level median or quartile interpolation over participant aggregates, not fractional events or route-level pseudo-replication. Questionnaire rows marked as grouped item scores are exploratory subjective indicators, not validated psychometric constructs.
\end{table*}

\subsubsection{Structured Route Access}

As summarized in Table~\ref{tab:results_summary}, \textit{HoldLight} yielded higher route-understanding scores than the controlled on-demand human-support condition. Descriptively, while baseline pre-climb previews often supported start/end identification, \textit{HoldLight} enabled 10 of the 12 participants to correctly articulate the overall route direction and key transitional segments consistently across all three of their assigned routes. Importantly, this reflects an improvement in acquiring structured, rubric-defined route features rather than a generalized enhancement of all spatial comprehension.

\subsubsection{Uncertainty Recovery and Route Continuity}

As summarized in Table~\ref{tab:results_summary}, \textit{HoldLight} was associated with fewer uncertainty breakdown episodes than the controlled on-demand human-support condition.

Clarification-dependent interruptions showed the same direction of improvement, while total task time did not differ detectably between conditions. This pattern suggests that the system's primary benefit was in maintaining action continuity and supporting recovery from next-hold uncertainty, rather than increasing raw speed.

\subsubsection{Immediate Human Re-Entry for Routine Route-Information Clarification}

As summarized in Table~\ref{tab:results_summary}, \textit{HoldLight} reduced immediate human re-entry for routine route-information clarification within the controlled workflow.

To avoid treating channel shift as reduced support need, we distinguished system-handled clarification from human route-information re-entry: the commands \textit{repeat} and \textit{detail} were treated as clarification handled within \textit{HoldLight}, whereas human route-information re-entry was counted only when the guide resumed route-information support for the immediate move.

This result should not be interpreted as evidence that participants needed less support overall, or that the system was superior to skilled human support. Some repetition and clarification that would have been directed to the human guide in the controlled on-demand condition could instead be handled through the system's repeat and more-detail actions in the \textit{HoldLight} condition. We therefore interpret this result together with route-understanding scores, uncertainty-breakdown outcomes, fallback logs, and qualitative accounts.

Across the 36 \textit{HoldLight} trials, human route-information re-entry occurred 14 times in total (0.39 events per trial), primarily in fallback or degraded-confidence situations. This count excludes system-handled repetition and detail requests that did not lead to human re-entry. Safety-only interventions remained separately coded and did not differ detectably between conditions ($p = .85$), reinforcing that the observed difference concerned routine route-information clarification rather than the underlying human safety boundary.

\subsubsection{Perceived Autonomy and Subjective Responses}

As summarized in Table~\ref{tab:results_summary}, post-condition ratings showed higher perceived autonomy under \textit{HoldLight}. This autonomy item should be interpreted primarily as in-climb action independence (e.g., the ability to recover from uncertainty without waiting for a human prompt) rather than full-workflow independence, given that the initial system setup still required high interaction effort or sighted assistance in some cases.

For the exploratory grouped item scores, Table~\ref{tab:results_summary} shows a modest advantage for in-climb cue usefulness and uncertainty recovery under \textit{HoldLight}, while trust, safety, and acceptability remained high and were not statistically distinguishable in this small sample. Specifically, on the individual item asking whether they would want the system available in future sessions, all 12 participants responded positively, with 8 selecting \textit{strongly agree} and the remaining 4 selecting \textit{agree}. Taken together with the lower rate of explicit human-help requests, these findings suggest that \textit{HoldLight} supported greater perceived autonomy while reducing immediate human re-entry for routine route-information clarification under the study conditions.

\subsubsection{Descriptive Vision-Status and Experience Observations}

Because the study included small subgroups by vision status (blind $N=6$; low vision $N=6$) and climbing experience ($<6$ months: $N=3$; 6--12 months: $N=3$; 1--3 years: $N=6$), we did not conduct inferential subgroup comparisons. We instead inspected participant-level outcomes descriptively to determine whether the overall condition pattern appeared to be confined to one subgroup. Directionally similar advantages for \textit{HoldLight} in route understanding and uncertainty recovery were observed among both blind and low-vision participants and across the three experience categories. However, clarification and fallback needs varied within every subgroup, and we did not observe a consistent subgroup-specific pattern that would support claims about differential effectiveness by vision status or climbing experience. These observations are descriptive only; adequately powered subgroup comparisons remain future work.

\subsubsection{Technical Reliability, Fallback Behavior, and System Boundaries}

To contextualize these interaction findings, it is critical to report where the technical substrate succeeded and where it clearly degraded. Prior to climbing, the system required manual route-revision events in 7 of the 36 initial mobile scans (19.4\%), primarily to correct ambient lighting glare on holds.

During the 36 formal \textit{HoldLight} trials, fallback was explicitly triggered 42 times. Of these events, 27 (64.3\%) were initiated by the user (primarily using the ``repeat'' or ``more detail'' actions), while 15 (35.7\%) were triggered automatically by the system due to low tracking confidence.

We observed two clear boundary conditions where \textit{HoldLight} clearly did not help. First, on route segments where holds of the same color were placed closely together, the system’s progress inference frequently failed to disambiguate the target. This resulted in delayed cues or the system rapidly cycling through adjacent holds, temporarily confusing the climber. Second, in three distinct trials, the system falsely anchored onto a hold belonging to an adjacent, visually similar route. This generated a clearly misleading directional cue, forcing the user to trigger a manual fallback by pausing and requesting human help, which imposed a recovery penalty. The 14 instances of human information re-entry reported above were almost entirely precipitated by the system encountering these dense clusters or intersections.

\subsubsection{Qualitative Explanatory Themes}

Qualitative materials helped explain these differences in three recurring ways.
First, participants' route-understanding responses and post-condition comments suggested that overview-first preview helped them retain route structure rather than memorize a long sequence of holds. As one Phase C participant (P7) described it, the preview helped them remember the route \textit{``as sections instead of a long list of holds.''} Second, short cues appeared easier to act on under bodily load. Participants contrasted compact prompts such as \textit{``right hand, two o'clock, far''} with longer descriptions or underspecified phrases such as \textit{``above you,''} which required additional searching while hanging on the wall. Third, fallback episodes showed that autonomy depended on graded recovery rather than full independence. As another Phase C participant (P10) put it, \textit{``It did not replace the guide, but it let me decide when I needed the guide.''} Consistent with this account, fallback logs and observation notes showed that human route-information re-entry remained necessary for dense same-color hold clusters, adjacent-route ambiguity, and safety-relevant uncertainty. These qualitative materials explain how the measured differences in route understanding, next-hold recovery, and perceived autonomy were experienced in practice, while remaining contextual evidence rather than separate causal proof.

\section{Discussion}

In this study, we proposed three linked questions about accessible route-information support for blind and low-vision climbers. In this section, we discuss the research findings, implications, and limitations. 

Regarding \textbf{RQ1}, our formative inquiry suggests that the central barrier is not simply lack of route information, but mismatch between how route information is commonly delivered and what climbers can actually use under bodily load. Route preview was often fragmented and difficult to retain, real-time prompting was useful but easy to overload, uncertainty required explicit recovery rather than hidden confidence, and participants repeatedly described effective support in terms of maintaining pace, exploration, and user control rather than maximizing instruction alone. In this sense, participants’ accounts suggest that effective support in climbing must do more than provide route content: it should help users form an initial route model, act on next-hold information when it becomes actionable, recover when guidance becomes uncertain, and reduce unnecessary dependence on repeated routine prompting.

Regarding \textbf{RQ2}, our design and pilot results suggest that, in this setting, accessible climbing guidance is best structured as a bounded interaction workflow combining layered pre-climb preview, concise action-timed cueing, explicit fallback, and bounded human collaboration. A core implication of this work therefore concerns the paradigm of explicit fallback. In our study, fallback was not treated as a system failure or a mere engineering backup, but as a first-class accessibility feature. When operating under physical load, a usable AI interaction depends less on maintaining an illusion of omniscient confidence than on making uncertainty legible so that users can better detect potentially unreliable guidance and decide how to recover \cite{Chen2025SurfacingVariations}. By supporting graded recovery, explicit fallback enhances autonomy by empowering users to remain in control of when and how assistance escalates, aligning with HCI work arguing that autonomy in assistive technology should not be reduced to independence alone \cite{Guldenpfennig2019AutonomyPerspective,Bennett2018InterdependenceFrame}. 

Regarding \textbf{RQ3a}, the controlled evaluation suggests that this workflow improved two specific aspects of access: structured route understanding before ascent and recovery from next-hold uncertainty during ascent. A critical nuance here concerns the value of structured pre-climb delivery. The layered, user-driven nature of \textit{HoldLight}'s route preview enabled climbers to build a more coherent mental model than the more constrained pre-climb summary used in the controlled on-demand human-support condition. This interpretation is consistent with broader accessibility work showing the value of context-sensitive and user-adjustable information delivery in domains beyond climbing \cite{Kuribayashi2025WanderGuide,Chang2024WorldScribe,Mohanbabu2024ContextAwareImageDescriptions}. At the same time, the system reduced uncertainty breakdown episodes and clarification-dependent interruptions without producing a clear task-time advantage. This pattern suggests that the main value of the workflow was not raw speed, but improved continuity of action: participants more often began the climb with a workable route model and were less likely to enter extended breakdown-repair cycles simply to recover from next-hold uncertainty.

Finally, regarding \textbf{RQ3b}, the difference between \textit{HoldLight} and the controlled on-demand human-support condition was not only procedural but also relational. Participants reported higher perceived autonomy under HoldLight, while explicit requests for human route-information help were substantially lower. Importantly, this should not be interpreted as full independence. As our results show, initial setup could still require substantial interaction effort or sighted assistance, and human information re-entry remained necessary in boundary cases such as dense same-color hold clusters and adjacent-route ambiguity. The autonomy supported here is therefore better understood as graded, in-climb control over when support is needed and how it escalates, rather than independence from others in any absolute sense. Our findings suggest that future embodied accessibility systems may be especially useful when they absorb part of ordinary information support while preserving human roles in ambiguity, safety, and high-context judgment, supporting forms of autonomy that depend on retained user control and bounded collaboration rather than full independence \cite{Guldenpfennig2019AutonomyPerspective,Bennett2018InterdependenceFrame}. More broadly, because BLV users' practical priorities do not always align closely with researcher assumptions or technical emphases \cite{Gamage2023AssistiveSmartDevices}, these findings reinforce the importance of grounding system behavior in user-valued forms of support rather than purely technical task framings.

% \section{Limitations}

This work has several limitations. First, the study used a modest sample in a controlled indoor setting. While appropriate for an initial accessibility evaluation, the results should not be interpreted as establishing broad generalizability across all BLV climbers, route styles, or climbing environments.
Second, our participant group included heterogeneity in visual condition and climbing background. This diversity was important for ecological relevance, but it also means that the system may not have worked equally well for all users or for all forms of uncertainty.
Third, although we enforced a minimum 10-minute rest period and used a counterbalanced condition order, indoor climbing is physically demanding. Residual physical fatigue and learning effects across the session remain potential confounders in our within-subjects design, limiting the purity of the in-climb continuity results.
Fourth, the comparison was bounded by design. We compared \textit{HoldLight} against a controlled on-demand human-support condition derived from current support practices, not against unconstrained expert coaching or unsupported climbing. Because \textit{HoldLight}'s layered, user-driven route preview inherently afforded more structured pre-climb interaction than the human guide's constrained summary, gains in route understanding should be interpreted as the benefit of this structured interaction workflow as a whole rather than the isolated effect of algorithmic route parsing itself. Likewise, because the guide followed a predefined prompt policy that may be more conservative than real-world experts reacting to subtle bodily cues, continuity-related gains should be interpreted relative to this bounded controlled comparison rather than as evidence of superiority over unconstrained expert human support.
Fifth, the technical substrate was sufficient to support the workflow studied here, but not uniformly reliable across all situations. As reported in the Results, route revision prior to interaction, dense hold clusters, and visually similar intersecting routes remained important boundary conditions for reliable performance. Furthermore, because our implementation relied on a server-assisted architecture rather than fully on-device processing, we did not conduct a formal end-to-end latency benchmark. Accordingly, delayed cues, retakes, and fallback events should be interpreted as workflow boundary observations rather than as a quantified latency evaluation. The system's benefits should therefore be interpreted in light of these technical boundaries.
% Sixth, because the study compared a system condition with a standardized human-guidance condition in a live climbing context, researchers and the baseline guide were aware of condition assignment. We used pre-specified operational definitions, condition-neutral coding rules, and video verification for ambiguous cases. However, we did not conduct formal inter-rater reliability analysis across all observation-derived events, meaning some interpretation bias may remain, particularly in qualitative accounts and borderline pausing events.
Finally, the strongly positive subjective ratings and positive short-term future-use responses must be interpreted cautiously. These outcomes were gathered after a short-term, highly controlled exposure, where the novelty effect---a common phenomenon in technology adoption and early system evaluation---may have elevated acceptability scores for a new accessibility prototype \cite{Long2024NotJustNovelty}. More broadly, this study did not evaluate longer-term adoption, training effects, or use in less controlled climbing contexts. Future work should examine how support preferences evolve over time, how these patterns differ across adequately powered subgroups of vision status and climbing experience, where the workflow clearly stops helping, and how structured route-information support interacts with richer real-world coaching and social climbing practices.

\section{Conclusion}

In this paper, we present \textit{HoldLight}, an accessibility system to be used before and during indoor rock climbing, aiming to support blind and low-vision climbers in understanding climbing routes and receiving effective guidance. Our work contributes a bounded accessibility workflow for indoor climbing that combines layered route preview, concise next-hold cueing, explicit fallback under uncertainty, and retained human safety collaboration. Across formative inquiry, pilot refinement, and a controlled initial evaluation, our findings suggest that this structure can improve route understanding, support recovery from next-hold uncertainty, and reduce immediate human re-entry for routine route-information clarification under controlled indoor conditions. At the same time, the results reinforce that such systems should be interpreted as evidence for a bounded accessibility workflow, not as evidence that automated support can replace flexible expert coaching or human safety judgment.

\begin{acks}
We thank all climbers, guides, coaches, and climbing-gym staff for their participation and support in this study.
\end{acks}

\bibliographystyle{ACM-Reference-Format}
\bibliography{references}

\newpage
\appendix

\section{Study Instruments}
\label{sec:appendix-studyinstruments}
To support methodological transparency while keeping the main paper focused on the core research questions and findings, we used a structured set of study instruments across the formative and controlled evaluation phases. The instruments were designed to capture participant background, current support practices, structured route access, uncertainty recovery during climbing, immediate human re-entry for routine route-information clarification, and post-task perceptions of autonomy, trust, safety boundary, and acceptability, together with qualitative reflections on human--system boundaries and dependence on human support.

\subsection{Phase A Instruments}
Phase A combined a short accessible background questionnaire with semi-structured interviews. The background questionnaire was used to characterize the participant pool and to sensitize later interviews rather than to serve as an outcome measure. It covered participants' visual condition, climbing experience, typical support arrangements, preferred information channels during movement, current barriers in climbing, and broad boundaries of acceptance for technology-assisted support.

The semi-structured interview for blind and low-vision (BLV) climbers focused on current route-preview practices, in-climb next-hold localization, the timing and wording of prompts, uncertainty handling, the distinction between information support and safety support, autonomy, enjoyment, and unacceptable support experiences. To move beyond general preferences, the interview also used critical-incident prompts to elicit concrete positive and negative episodes from prior climbing experiences, helping participants move beyond abstract preferences and describe specific moments of successful support, breakdown, and recovery \cite{Flanagan1954CriticalIncidentTechnique}.

In addition, we conducted a separate semi-structured interview with secondary stakeholders, namely sighted guides and coaches with relevant experience supporting blind and low-vision climbers. This interview was not intended to replace the perspectives of BLV climbers. Instead, it was used to understand the current support ecology, common communication breakdowns, supporter workload, candidate elements of a controlled on-demand human-support condition derived from current support practices, and the appropriate boundary between technological assistance and human judgment.

\subsection{Phase C Instruments}
For the controlled initial evaluation, each formal condition was followed by a single-condition post-test questionnaire. All closed-ended questionnaire items used a 5-point agreement scale from 1 (\textit{strongly disagree}) to 5 (\textit{strongly agree}). Table~\ref{tab:questionnaire_items} lists the questionnaire items and the exploratory grouped item scores used for descriptive and inferential analysis. These grouped scores were not treated as validated psychometric constructs. The questionnaire was organized primarily around two exploratory grouped item scores---in-climb cue usefulness and uncertainty recovery, and trust, safety boundary, and acceptability---plus one single-item autonomy outcome. Items related to perceived route access were retained as descriptive and interpretive support alongside the rubric-based route-understanding task rather than treated as a separate primary questionnaire construct in the main inferential results. Reduced reliance on ordinary human information support was not treated as a separate Likert construct, because the questionnaire did not include a direct multi-item cluster for that concept. Instead, immediate human re-entry for routine route-information clarification was interpreted from behavioral measures, open-ended responses, and the final comparative interview. Open-ended items were included to capture the most helpful aspect of the condition, the most difficult or disruptive aspect, the participant's highest-priority improvement, and which tasks still appeared better suited to human support.

Table~\ref{tab:route_understanding_rubric} presents the scoring rubric for the pre-climb route-understanding task.

\begin{table*}[h]
\caption{Scoring rubric for the pre-climb route-understanding task. Total score range: 0--4.}
\label{tab:route_understanding_rubric}
\centering
\small
\setlength{\tabcolsep}{3pt}
\renewcommand{\arraystretch}{1.05}
\tagpdfsetup{table/header-rows={1}}
\begin{tabular}{@{}p{0.18\linewidth} p{0.23\linewidth} p{0.23\linewidth} p{0.23\linewidth} p{0.07\linewidth}@{}}

\toprule
\textbf{Scored feature} & \textbf{Correct (1.0)} & \textbf{Partially correct (0.5)} & \textbf{Incorrect (0.0)} & \textbf{Max} \\
\midrule
Start identification & Correctly identifies the target route's starting holds or start region. & Identifies an approximately correct start area but not the correct holds or not with sufficient specificity. & Fails to identify the start or identifies a clearly incorrect start. & 1.0 \\
\addlinespace
End identification & Correctly identifies the target route's finishing holds or end region. & Identifies an approximately correct end area but not the correct finish or not with sufficient specificity. & Fails to identify the end or identifies a clearly incorrect end. & 1.0 \\
\addlinespace
Overall route direction / major transition & Correctly describes the general route trajectory and at least one major directional change or key transition. & Provides only a partially correct trajectory, a vague upward description, or an incomplete account of the route's main turn. & Cannot describe the route trajectory in a usable way or describes a clearly incorrect path. & 1.0 \\
\addlinespace
Key segment / key move & Correctly identifies a salient segment, key transition, difficult section, or likely important move on the route. & Mentions a vaguely relevant area without clearly identifying a meaningful key segment or move. & Does not identify any meaningful key segment / move, or identifies one that is clearly unrelated to the target route. & 1.0 \\
\midrule
{} & {} & {} & \hfill\textbf{Total possible score} & \textbf{4.0} \\
\bottomrule
\end{tabular}
\end{table*}

After both conditions were completed, participants took part in a final semi-structured comparative interview. This interview first reviewed each condition separately and only then moved to explicit cross-condition comparison. It focused on route understanding before climbing, usefulness and timing of in-climb guidance, rhythm and cognitive burden, uncertainty handling, trust, autonomy, social dependence, human--system boundary, unacceptable experiences, and improvement priorities.

During each formal task, researchers additionally completed two structured records. First, a structured observation form was used to document route-understanding behavior, visible pauses, clarification episodes, rhythm disruption, uncertainty, help-seeking, and critical incidents. Second, an objective metrics sheet was used to capture time-based and count-based measures such as route-understanding task time, total task time, time to first major pause, cumulative pause duration, repetition and clarification requests, extra human information-support intervention, next-hold localization failures, wrong-route deviations, system-to-human fallback events where applicable, and prompt-quality issues such as delay, error, or ambiguity.

\section{Phase C Questionnaire Grouped Item Scores and Items}
\label{sec:appendix-questionnaire}
All closed-ended questionnaire items used a 5-point agreement scale from 1 (\textit{strongly disagree}) to 5 (\textit{strongly agree}).
Table~\ref{tab:questionnaire_items} lists the post-condition questionnaire items used in Phase C.

\begin{table*}[h]
\caption{Post-condition questionnaire items used in Phase~C, organized by analytic outcome grouping. All closed-ended items used a 5-point agreement scale from 1 (\textit{strongly disagree}) to 5 (\textit{strongly agree}).}
\label{tab:questionnaire_items}
\centering
\small
\tagpdfsetup{table/header-rows={1}}
\begin{tabular}{p{0.22\linewidth} p{0.08\linewidth} p{0.62\linewidth}}
\toprule
\textbf{Outcome grouping} & \textbf{Item ID} & \textbf{Item wording} \\
\midrule

\textbf{Structured route access} & & \\
& Q1 & Before starting the climb, this method helped me better understand the whole route. \\
& Q2 & Before starting the climb, I was able to form a relatively clear overall picture of the route. \\
\midrule

\textbf{In-climb cue usefulness and uncertainty recovery} & & \\
& Q3 & During climbing, this method helped me better find the next hold I should move to. \\
& Q4 & Overall, the information or prompts I received were clear and easy to understand. \\
& Q5 & Overall, the timing of the information or prompts was appropriate. \\
& Q6 & Overall, the amount of information was appropriate; it did not feel too much, too confusing, or insufficient. \\
& Q7 & This method did not noticeably interrupt my climbing rhythm. \\
& Q8 & I was able to maintain good focus while using this method. \\
& Q9 & When I was uncertain about the next move, this method generally helped me know what to do next. \\
& Q10 & When I did not hear clearly, did not understand, or the information felt insufficient, I knew how to request repetition, clarification, pause, or help. \\
\midrule

\textbf{Perceived autonomy} & & \\
& Q11 & This method made me feel more autonomous while climbing. \\
\midrule

\textbf{Trust, safety boundary, and acceptability} & & \\
& Q12 & Overall, I felt relatively safe and at ease during this trial. \\
& Q13 & Overall, I was willing to trust the information or support provided by this method. \\
& Q14 & This method could be integrated into climbing naturally, rather than creating additional hassle. \\
& Q15 & Overall, this method was helpful to me. \\
& Q16 & Overall, this method helped preserve my enjoyment of climbing. \\
& Q17 & If this method were available in the future, I would be willing to use it again in suitable situations. \\
\midrule

\textbf{Open-ended items} & & \\
& Q18 & What was the most helpful part of this experience? \\
& Q19 & What was the most difficult, confusing, or experience-disrupting part of this experience? \\
& Q20 & If only one thing could be improved first, what should it be? \\
& Q21 & In this experience, which tasks still seemed better suited to human support? \\
\bottomrule
\end{tabular}
\end{table*}

\section{Scoring and Coding Summary}
\label{sec:appendix-scoring}
To support transparent interpretation of the controlled evaluation, we defined a small set of structured scoring and coding procedures before formal data collection. These procedures were used to score pre-climb route understanding and to record observable interaction breakdowns, recovery actions, and immediate human re-entry for routine route-information clarification.

\subsection{Operational Definitions for In-Climb Event Coding}

To improve coding consistency, the research team used a shared set of operational definitions before formal data collection. Table~\ref{tab:event_definitions} summarizes the core event definitions used in Phase C coding.

\begin{table*}[h]
\caption{Operational definitions used for in-climb event coding and structured observation.}
\label{tab:event_definitions}
\centering
\small
\tagpdfsetup{table/header-rows={1}}
\begin{tabular}{p{0.28\linewidth} p{0.64\linewidth}}
\toprule
\textbf{Event / Code} & \textbf{Operational Definition} \\
\midrule
\textbf{Major Pause / Uncertainty-Driven Breakdown}
& A cessation of upward vertical progression lasting $\geq 6$ seconds. To be coded as an uncertainty-driven breakdown, the pause also had to be accompanied by exploratory searching without committed weighting, explicit verbalization of uncertainty, or a clear pause for route-related decision-making or guidance-seeking. Pauses explicitly requested for physical rest were not coded as uncertainty-driven breakdowns. \\
\addlinespace
\textbf{Exploratory Touch}
& Tapping, sweeping, or briefly touching a non-target hold to gather spatial information without shifting meaningful body weight onto it for upward progression. Exploratory touch was not automatically coded as an error. \\
\addlinespace
\textbf{Wrong-Hold Deviation}
& The participant clearly grasped or weighted a non-target hold in a way that functionally contributed to upward progression. \\
\addlinespace
\textbf{Wrong-Route Deviation}
& The participant's action sequence clearly diverged from the intended route path rather than reflecting brief exploratory contact alone. \\
\addlinespace
\textbf{Clarification-Dependent Interruption}
& The participant explicitly halted or suspended ongoing movement in order to request repetition, clarification, confirmation, or more detail about the next action. \\
\addlinespace
\textbf{System-to-Human Fallback Event}
& The system shifted into a conservative or degraded support state and route-information support for the immediate move was subsequently provided by a human guide, either following a system-triggered fallback or a user-triggered request for human help. \\
\addlinespace
\textbf{Delayed Prompt}
& The participant had clearly entered a next-hold search state, but the prompt arrived late enough to produce an observable stall in momentum under the agreed coding rules. \\
\addlinespace
\textbf{Prompt Error / Prompt Ambiguity}
& \textit{Prompt error:} guidance clearly indicated an incorrect target, direction, or route segment. \newline
\textit{Prompt ambiguity:} the participant heard the prompt but the wording was too unclear or underspecified to act on reliably. \\
\addlinespace
\textbf{Safety-Only vs.\ Information-Support Intervention}
& Safety-only interventions were coded separately from information-support interventions wherever possible so that retained human safety oversight would not be conflated with ordinary route-information support. \\
\bottomrule
\end{tabular}
\end{table*}

\subsection{Observation and Qualitative Analysis}
The structured observation form was used primarily as contextual and explanatory evidence. Researchers first recorded concrete behavior and salient incidents and only then added higher-level interpretations when needed. Observation categories included pre-climb route-understanding behavior, visible uncertainty, pauses, clarification episodes, rhythm disruption, moments of extra help-seeking, and safety-related escalation.

Because live-condition observation can be influenced by condition awareness, researchers used pre-specified operational definitions and condition-neutral coding rules. Video-based verification was used for ambiguous cases where available. These qualitative and observational materials were used to explain the mechanisms behind quantitative patterns and to identify success criteria and unacceptable experiences beyond task completion alone.

Because observation was used primarily as contextual and explanatory evidence, we report these core rules for transparency but did not conduct a full formal inter-rater reliability analysis across all observation-derived events.

\section{Deployed Runtime Architecture}
\label{sec:appendix-runtimearchi}
\begin{figure*}[t]
  \centering
  \includegraphics[
  width=\textwidth,
  alt={A technical runtime architecture diagram for the deployed HoldLight prototype. The diagram shows users, a browser frontend, browser-local capabilities, reverse proxy or static hosting, a Node and Express application programming interface, a worker pool, a Python vision runtime, and MongoDB storage. It also shows two pipelines: wall scan and live guidance, including browser-local pose tracking, server-assisted inference and calibration, frontend route planning, and spoken audio-cue output.}
  ]{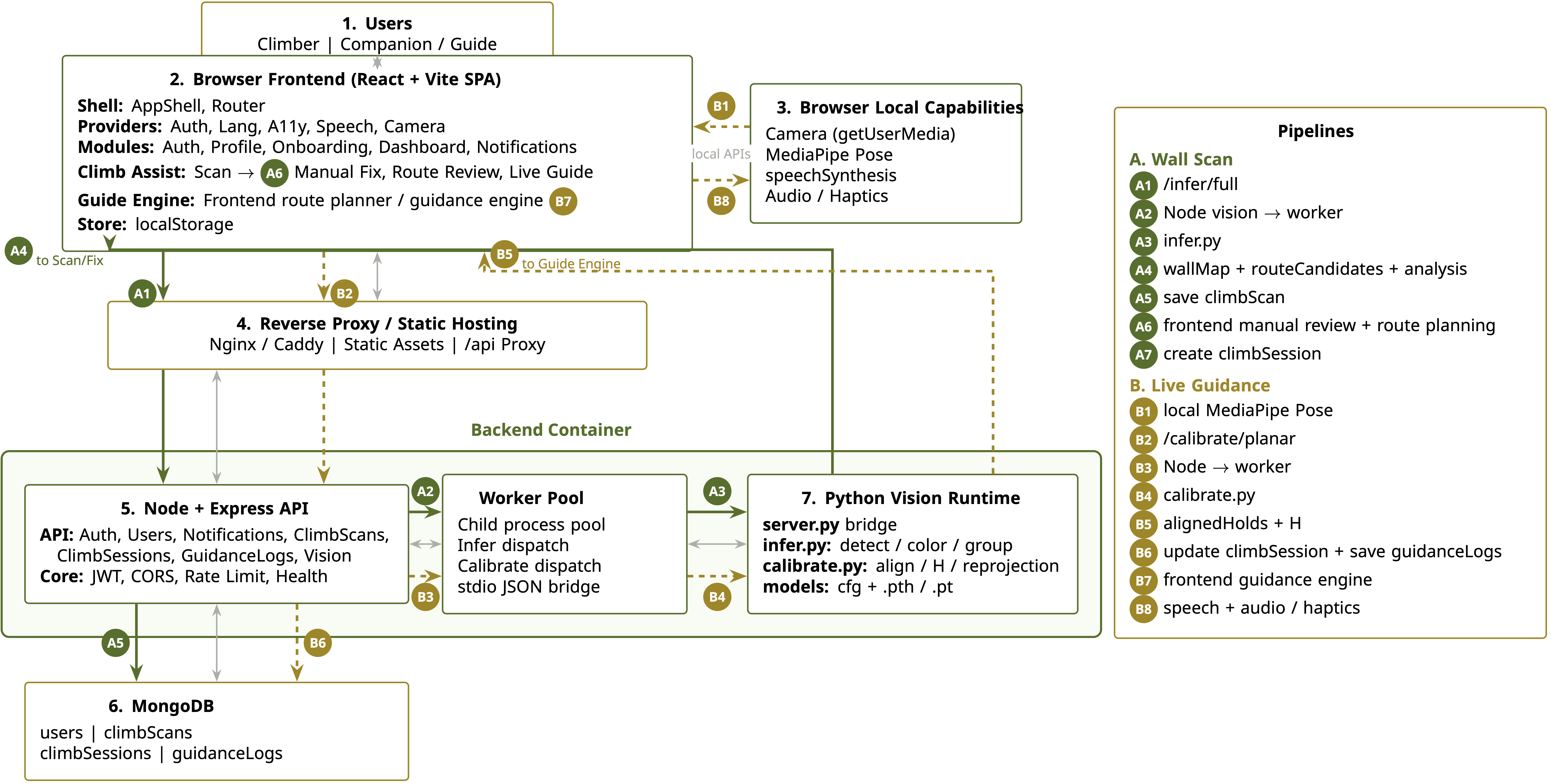}
  \caption{Deployed runtime architecture of the HoldLight prototype. The figure details the browser-centric implementation stack, including the React/Vite frontend, browser-local capabilities, reverse proxy/static hosting, Node/Express API layer, worker pool, Python vision runtime, and MongoDB persistence, together with the main wall-scan and live-guidance execution pipelines.}
  \Description{A technical runtime architecture diagram for the deployed HoldLight prototype. The diagram shows users, a browser frontend, browser-local capabilities, reverse proxy or static hosting, a Node and Express application programming interface, a worker pool, a Python vision runtime, and MongoDB storage. It also shows two pipelines: wall scan and live guidance, including browser-local pose tracking, server-assisted inference and calibration, frontend route planning, and spoken audio-cue output.}
  \label{fig:appendix-runtime-architecture}
\end{figure*}

Figure~\ref{fig:appendix-runtime-architecture} provides a lower-level view of the deployed prototype architecture and execution flow that complements the workflow-centered overview in Fig.~\ref{fig:system-overview}. It shows the browser-centric implementation, the division between browser-local capabilities and server-assisted processing, and the two main pipelines used in the prototype: wall scan and live guidance.

\end{document}